\documentclass[]{aastex631}
\usepackage{amsmath}[reqno]
\usepackage{appendix}

\newcommand{\framework}{\textit{NExSS Quantitative Habitability Framework}}

\newcommand{\celsius}{$^\circ$C}

\shorttitle{NExSS Framework for Habitability}
\shortauthors{QuantHab SWG}
\graphicspath{{./}{}}

\newcommand{\qfw}{\textsc{NExSS Quantitative Framework for Habitability}}

\begin{document}

\newcommand{\TRe}{TRAPPIST-1e-like}
\newcommand{\TRf}{TRAPPIST-1f-like}

\title{A Terminology and Quantitative Framework for Assessing the Habitability of Solar System and Extraterrestrial Worlds}

\correspondingauthor{D\'aniel Apai}
\email{apai@arizona.edu}
\author[0000-0003-3714-5855]{D\'aniel Apai}
\affil{Steward Observatory, The University of Arizona, 933 N. Cherry Avenue, Tucson, AZ 85721, USA}
\affiliation{Lunar and Planetary Laboratory, University of Arizona, 1629 E. University Boulevard, Tucson, AZ 85721, USA}
\affiliation{Alien Earths Team, NASA ICAR/NExSS, USA}

\author[0000-0001-6487-5445]{Rory Barnes}
\affiliation{Astronomy Department, University of Washington, Box 951580, Seattle, WA 98195, USA}

\author[0000-0002-8517-8857]{Matthew M. Murphy}
\affil{Steward Observatory, The University of Arizona, 933 N. Cherry Avenue, Tucson, AZ 85721, USA}
\affiliation{Alien Earths Team, NASA ICAR/NExSS, USA}

\author[0000-0002-3286-7683]{Tim Lichtenberg}
\affil{Kapteyn Astronomical Institute, University of Groningen, P.O. Box 800, NL-9700 AV Groningen, The Netherlands}
\affiliation{Alien Earths Team, NASA ICAR/NExSS, USA}

\author[0000-0003-3989-5545]{Noah Tuchow}
\affil{NASA Goddard Space Flight Center, Greenbelt, Maryland, USA}

\author[0000-0002-5806-5566]{R\'egis Ferri\`ere}
\affil{Institut de Biologie de Ecole Normale Superieure École Normale Supérieure, Université Paris Sciences et Lettres, 46 Rue d’Ulm, F-75005 Paris, France}
\affil{International Research Laboratory for Interdisciplinary Global Environmental Studies (iGLOBES) CNRS, ENS, Universit\'e Paris Sciences et Lettres, University of Arizona, Tucson, AZ, USA}
\affil{Department of Ecology and Evolutionary Biology, University of Arizona, Tucson, AZ, USA}
\affiliation{Alien Earths Team, NASA ICAR/NExSS, USA}

\author[0000-0002-4309-6343]{Kevin Wagner}
\affiliation{Department of Astronomy and Steward Observatory, University of Arizona, USA}

\author[0000-0003-3481-0952]{Antonin Affholder}
\affil{Department of Ecology and Evolutionary Biology, University of Arizona, Tucson AZ, USA.}
\affiliation{Alien Earths Team, NASA ICAR/NExSS, USA}

\author[0000-0002-1226-3305]{Renu Malhotra}
\affiliation{Lunar and Planetary Laboratory, University of Arizona, 1629 E. University Boulevard, Tucson, AZ 85721, USA}
\affiliation{Alien Earths Team, NASA ICAR/NExSS, USA}

\author[0000-0002-0957-3177]{Baptiste Journaux}
\affiliation{Department of Earth and Space Science, University of Washington, Box 951310, Seattle, WA 98195, USA}

\author[0000-0001-9504-3174]{Allona Vazan}
    \affiliation{Astrophysics Research Center (ARCO), Department of Natural Sciences, The Open University of Israel, Raanana 4353701, Israel}

\author[0000-0001-7553-8444]{Ramses Ramirez}
\affiliation{University of Central Florida, Department of Physics, Planetary Sciences Group, Orlando, Fl. 32816}

\author[0000-0003-0726-0748]{Abel M\'endez}
\affil{Planetary Habitability Laboratory, University of Puerto Rico at Arecibo (abel.mendez@upr.edu)}

\author[0000-0002-7084-0529]{Stephen R. Kane}
\affil{Department of Earth and Planetary Sciences, University of California, Riverside, CA 92521, USA}

\author{Veronica H. Klawender}
\affil{Steward Observatory, The University of Arizona, 933 N. Cherry Avenue, Tucson, AZ 85721, USA}

\author{NExSS Quantitative Habitability Science Working Group}

 \newcommand{\Hq}{$H_{\mathrm q}$}

\begin{abstract}

The search for extraterrestrial life in the Solar System and beyond is a key science driver in astrobiology, planetary science, and astrophysics. A critical step is the identification and characterization of potential habitats, both to guide the search and to interpret its results. However, a well-accepted, self-consistent, flexible, and quantitative terminology and method of assessment of habitability are lacking.
Our paper fills this gap based on a three year-long study by the NExSS Quantitative Habitability Science Working Group. We reviewed past studies of habitability, but find that the lack of a universally valid definition of life prohibits a universally applicable definition of habitability. A more nuanced approach is needed.
We introduce a quantitative habitability assessment framework (QHF) that enables self-consistent, probabilistic assessment of the compatibility of two models: First, a habitat model, which describes the probability distributions of key conditions in the habitat. Second, a viability model, which describes the probability that a metabolism is viable given a set of environmental conditions.
We provide an open-source implementation of this framework and four examples as a proof of concept: (a) Comparison of two exoplanets for observational target prioritization; (b) Interpretation of atmospheric O$_2$ detection in two exoplanets; (c) Subsurface habitability of Mars; and (d) Ocean habitability in Europa. These examples demonstrate that our framework can self-consistently inform astrobiology research over a broad range of questions. The proposed framework is modular so that future work can expand the range and complexity of models available, both for habitats and for metabolisms.
\end{abstract}

\keywords{}


\section{Introduction}
\label{S:Introduction}

\begin{figure}[ht!]
\begin{center}
\includegraphics[width=0.94\linewidth,angle=0]{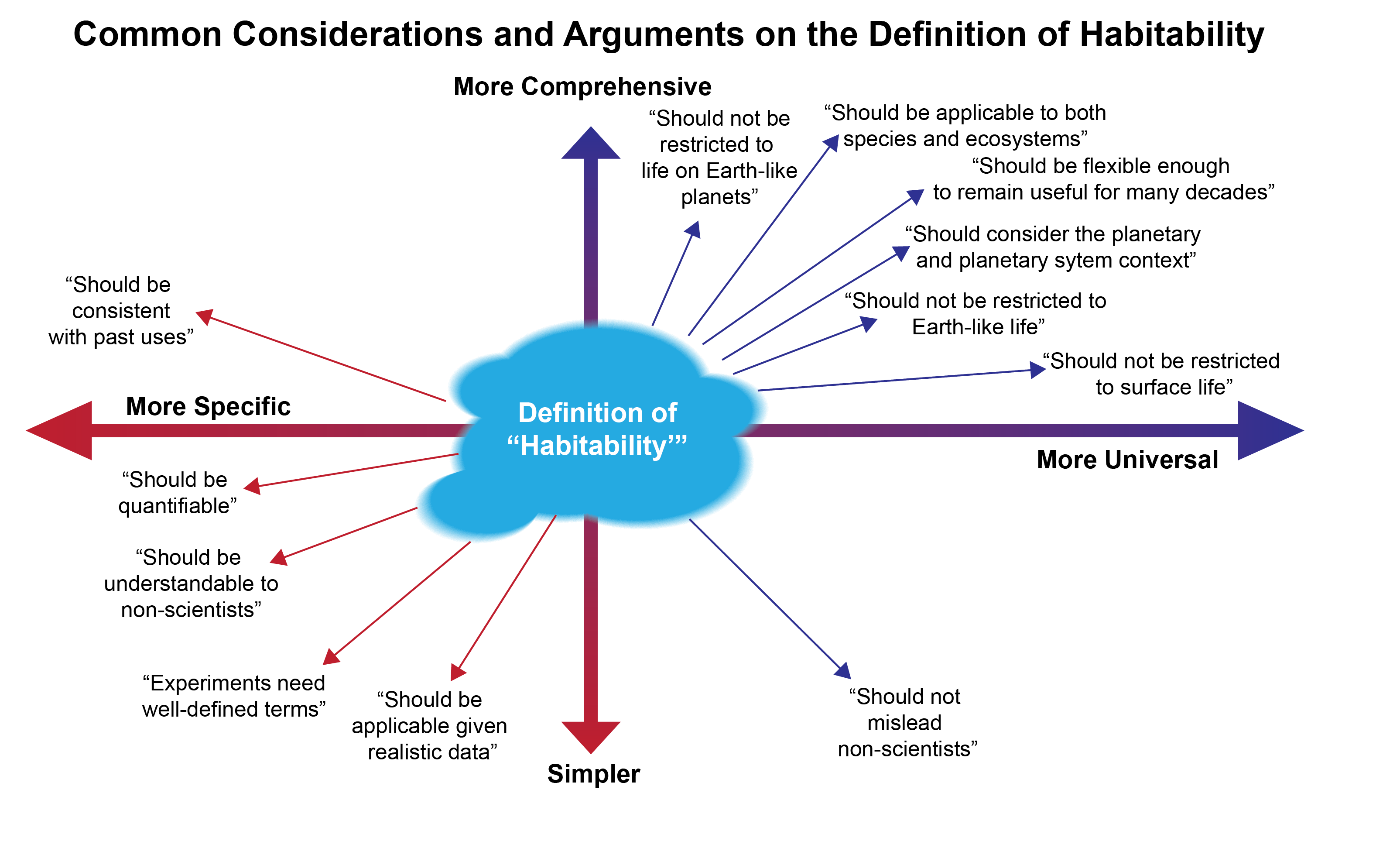}
\caption{Visual summary of key considerations and arguments about the nature and scope of the definition and use of the term ``habitable''. There are strong and valid arguments both for a simpler and more specific meaning and for a more complex and more universal meaning. Any single {definition} for ``habitability'' fails to meet the majority of the requirements. \label{f:Considerations}}
\end{center}
\end{figure}

With increasing capabilities in in-situ robotic probes and remote-sensing observatories, humanity may soon be within reach of identifying life beyond Earth. The search for life has emerged as one of the most important science drivers in astronomy (e.g., \citealt[][]{Astro2020}), planetary sciences, and astrobiology (e.g., \citealt[][]{PlanetaryDecadal2023}). Integral to that search is the study and identification of potential habitats suitable for life: From the Martian sub-surface to the clouds of Venus, from the oceans of dwarf planets and icy moons to extrasolar worlds. The term ``habitable'' (and its derivative ''habitability'') are widely used in our disciplines and even appear in the highest levels of mission goals (e.g., NASA's ``Habitable Worlds Observatory'', ESA's to-be-selected L5 mission), yet their exact meaning and quantitative interpretation differs widely between investigations. In fact, a review of the literature reveals significant tension in the use of these words: Compelling arguments are made for more specific vs. more universal, and simpler vs. more comprehensive definitions of habitability and related terms (see Figure~\ref{f:Considerations} for a visual summary).
Unsurprisingly, the popular media (and occasionally research investigations) also often confuse ``habitability'' with the ``habitable zone'', in spite of important differences between them.

We argue that -- in order to increase robustness and falsifiability of future research associated with these large-scale science questions -- \textit{it is very important for the fields of astrobiology, planetary sciences, and astronomy that a clear, quantitative, expandable, flexible, and widely applicable terminology is introduced, and that it is widely adopted by
the community}. This paper presents a framework developed to address these needs. Our study is the result of a three-year study conducted by the NExSS Quantitative Habitability Science Working Group (chairs: D. Apai and R. Barnes). The SWG held regular meetings between 2020 and 2023, solicited input from members of the NExSS network and from the broader community, held a workshop (with over 100 participants), and developed the present paper through an iterative process with the community.

This paper is organized as follows: In Section~\ref{S:TerminologyConsiderations}, we provide the necessary philosophical and scientific background for the definition of a new scientific terminology for habitability assessment. In Section~\ref{S:CommunityReports}, we review the strategic community reports on exoplanet habitability and astrobiology with special focus on their vision and handling of habitability. In Section~\ref{S:SurveyTerminology}, we survey the research literature for relevant terminology and illustrative quantitative frameworks. In Section~\ref{S:NewFramework}, we introduce a new terminology and, building on that, the \framework. In Section~\ref{S:V-functions}, we provide examples for viability functions, followed by four example applications in Section~\ref{S:Applications}. In Section~\ref{S:Discussion}, we discuss the key properties, applications, and limitations of the new terminology and framework. In Appendix~\ref{S:Appendix}, we provide an overview of the open source python implementation of the \framework{} introduced in our study, complete with four examples. Finally, in Section~\ref{S:Summary}, we summarize the key findings of our study.

\section{Considerations for Scientific Terminology}
\label{S:TerminologyConsiderations}

\subsection{The Problem at the Foundations of Habitability} \label{S:DefinitionOfLife}

Arguably, it is important to identify at the onset of our study the fundamental problem that makes a perfect and ultimate definition of habitability challenging, if not impossible: \textit{Ultimately, the root of the problem that hampers the community's efforts to converge on a definition for habitability is that habitability depends on the requirements for life, and we do not have a widely agreed-upon definition for life.}

Much has been written and argued about universally required and sufficient properties of life, 
yet a definition, even after centuries of debate, remains elusive.
Some scholars argue that a search for a definition of life, at least with our present knowledge is doomed to fail but it may not even be necessary for continued progress \citep[e.g.,][]{Cleland2012}. One particularly insightful analogy may be the centuries of failed effort to define ``water'' that preceded the discovery of chemical elements and the existence of molecules. In this view, we may just lack a deeper level of understanding of living organisms that would enable us to capture their true essence -- how they differ from non-living objects.

Others argue that the definition of life may be a fool's errand regardless of a potential future, deeper understanding of life. For example, \citet{Machery2012} argues that life is practically undefinable: For many, life is a ``folk'' concept, i.e., an intuitive concept that feels well-defined but, when examined more rigorously, lacks a consistent basis. Thus, those attempting to define life often resort to coming up with criteria to support their intuition, rather then drawing from scientific observations. \citet{Machery2012} also argues that a scientific definition of life is likely impossible to reach, as life crosses multiple disciplines (e.g., evolutionary biology, astrobiology, biochemistry), and scholars of different disciplines experience (and therefore define) life differently, from fundamentally incompatible perspectives. Therefore, the author argues, life cannot be defined and attempts to do it should be abandoned.

\citet[][]{Smith2022}, in their commentary entitled ``The futility of exoplanet biosignatures'', takes this argument further: They argue that the lack of definition of life means that currently envisioned missions to search for extraterrestrial life are \textit{not testing well-defined hypotheses} but rather are searches in the dark. The authors argue that without a clear definition of life (or, well-defined hypotheses about living planets), the costs of biosignature search missions cannot be justified.

\subsection{Properties of Good Scientific Terminology Systems}

Scientific terminology -- the system of terms -- is an essential tool we use to describe and solve problems. During the history of science, terminology for problems has been introduced and refined. While good terminology systems advance the scientific community's ability to solve problems, and advance and communicate knowledge, suboptimal terminology can propagate confusion, misconceptions, or hide connections, thereby being detrimental to the field.
The lifetime of defined terms is not to be underestimated: Misleading terminology often remains in use for decades or even centuries. A few such examples include: the ``nebulae'' category that included, for example, galaxies; and ``planetary nebulae'', which represent the end stages of stellar evolution rather than the formation of planets, as it was originally believed.

We identify the following requirements as key to a \textit{successful} terminology: (A) The terminology is useful for addressing the phenomenon it aims to describe (no misleading terms); (B) The terminology has a long-term impact on the field: Oftentimes terms do not change for centuries; (C) It provides a  granularity that is practical for everyday as well as scientific use; (D) It should consist of a self-consistent {and substantially not overlapping} system of terms;
(E) All individual terms should be clearly defined;  (F) Conclusive, non-trivial statements can be made within the system of terms; (G) It has an {enduring}  legacy value: {meanings, definitions, and} conclusions are not expected to change substantially with more data.

\subsection{Pitfalls and Limitations}

Scientific nomenclature and terminology may fail to reach their intended goals for a number of reasons. Some past examples of less-than-ideal terminology systems provide useful insights into potential pitfalls. In the following, we will briefly summarize the key pitfalls that we identified as relevant in the context of quantifying habitability.

(A) Terminology not adopted widely: A terminology can only be truly useful if it is widely adopted by the community. If the terminology fails to gain support, it will -- regardless of its intrinsic merits -- not serve its purpose.

(B) Inconsistent (Self-Contradictory) Terminology: A system of terms that is, by definition, not self-consistent or carries within it contradictions, will inevitably lead to confusion in the field. For example, Population I stars are younger than Population II stars, which are in turn younger than Population III stars. Dark matter and dark energy are not simply ``dark'' but invisible. Brown dwarfs are not brown; or, a ``protoplanetary nebula'' is not related to protoplanets.
Beyond astronomy, misleading names are also common in biology (e.g., flying fox, jelly fish, tasmanian tiger, seahorse).
Some of these contradictory terms and terminology survive and remain in use in spite of the flawed nature of the underpinning assumptions that led to their emergence. Others -- perhaps the most incorrect ones -- have been replaced by efforts to bring clarity to the field. {For example, the term archeabacteria was introduced to refer to what was thought to be an ancient sub-group of bacteria -- but archaea reflect a new domain of life and not a subgroup of bacteria. Similarly, the term ``blue-green algae'' are still in use even though it refers to cyanobacteria, which are not algae (eukaryotes).\footnote{We thank the referee for pointing out these examples.}}

(C) Terminology is markedly different from everyday use of terms, which may mislead the public and create confusion. A good example of this pitfall may be the term ``theory'': In  everyday usage, it refers to a speculative idea without much supporting evidence. In contrast, in scientific terminology ``theory'' refers to the highest level of theoretical constructs or models, supported by a vast amount of evidence. Another pertinent example is the ``habitable zone'': We cannot blame non-specialists (or even journalists) for assuming that planets in the habitable zone should be habitable and that planets outside the habitable zone should not be able to harbor life. Yet, neither is true: The term habitable zone only expresses the potential of the right type of planets to harbor liquid water on its surface.

(D) The classification of important results based on terminology remain uncertain and \textit{requires frequent revisions}, potentially undermining the credibility of the field. As terminologies are, in part, attempts to map our models and classification to reality, living and evolving science should often necessitate the revision or adjustment of terminology. However, a frequently changing terminology poses a significant challenge to transferring information within scholars (e.g., from one decade to the next), and communicating results to non-specialists. Too rigid terminology systems based on too strictly-defined terms are unlikely to age well. Therefore, a successful terminology must retain enough flexibility that it can be adjusted as our understanding of the relevant phenomena deepens.

(E) Terminology built on definitions or criteria that are \textit{not applicable in practice} (e.g., requires knowledge of parameters that cannot be observed/derived). This issue is a particularly important threat to exoplanet science, which is currently ``data poor'': Inferences may need to be made about planets and potential alien ecosystems based on just a handful of parameters{, often with large uncertainties}. Terminology systems that would require the information content far exceeding the capabilities of instrumentation and surveys of the next decades would not be useful for the community.

In the following two sections, we will review recent uses of habitability in strategic reports and terminology, some of which are negatively impacted by some of the issues discussed above.  In Section~\ref{S:NewFramework}, we introduce a new framework which was created to avoid the five common problems identified above.

\section{Habitability in Strategic Reports and Concept Studies}
\label{S:CommunityReports}

The long-term strategy for exoplanet exploration and for the search of life on extrasolar planets is profoundly shaped by a few strategic reports commissioned by the United States' National Science Foundation (NSF), the National Academy of Arts and Sciences (NAS), and other international bodies, such as the European Space Agency in their Voyage 2050 report. It is, therefore, particularly important to understand the terminology used by recent, key reports and the methodology they endorsed for assessing planetary habitability. In the following, we briefly review the key aspects of four such reports: The NSF--NASA--DOE Exoplanet Task Force Report, the NAS Report on Exoplanet Strategy, NAS Report on Astrobiology Strategy, the HabEx and LUVOIR Science and Technology Definition Teams' reports, and the Astronomy \& Astrophysics 2020 Decadal Survey ``Pathways to Discovery'' Report.

\subsection{The NSF-NASA-DOES Exoplanet Task Force Report}
\label{S:NSFExoplanetTaskForce}

The NSF-NASA-DOE Astronomy and Astrophysics Advisory Committee (AAAC) established the ExoPlanet Task Force (ExoPTF) in 2006. Its charge was to advise NSF and NASA ``on the future of the ground-based and space-based search for and study of exoplanets, planetary systems, Earth-like planets and habitable environments around other stars''. The ExoPTF was asked to recommend a 15-year strategy to detect and characterize exo-planets and planetary systems, and their formation and evolution, including specifically the identification of nearby candidate Earth-like planets and study of their habitability\footnote{\url{https://www.nsf.gov/mps/ast/aaac/exoplanet_task_force/exoptf_charge_letter/exoptf_charge_final_signed.pdf}}. The ExoPTF report\footnote{\url{https://www.nsf.gov/mps/ast/aaac/exoplanet_task_force/reports/exoptf_final_report.pdf}} was submitted in 2008.

Not surprisingly, advancing the search for and understanding of habitable planets and planetary systems was a key component the report: Habitability is mentioned nearly 200 times in the report. However, the vast majority of these mentions focus on the habitable zone and the report refrains from recommending any quantitative assessment of habitability. Its Section 4.1.1 \textit{Defining Habitability and a Habitable Zone} states: ``We assume that life is carbon-based and that it requires at least the transient presence of liquid water. We are, of course, cognizant of other possibilities, but since we must focus our initial search efforts, it makes sense to begin with this assumption.'' Its Appendix 11.1 states that ``Habitability'' provides an essentially identical definition (emphasis added): \textit{``we further restrict our attention to {planets on which liquid water is present at the planet’s surface}. Surface life can alter the composition of a planet’s atmosphere in detectable ways, especially if it is powered by photosynthesis, as it is on Earth. The region around a star in which surface liquid water is stable is termed the habitable zone (HZ)''}, and derives from this requirement the habitable zone. It does mention, however, potential challenges unique to certain types of planets (e.g., planets around M-dwarf stars), suggesting that the definition introduced should not be considered universal:
\textit{``Only by observation will we learn whether any of them are actually inhabited.''}
With respect to shorter-term, practical approaches to assessing candidates for habitable planets, the report concludes:
``Temperature and atmospheric composition are critical for identifying habitable planets.''

In short, the report identifies habitable with the presence of surface water (as an admittedly imperfect starting point), but does not offer a quantitative strategy for assessing habitability. Instead, it recommends, essentially and implicitly (see \S\ref{S:Elimination}), the process of elimination as practical process to narrow down on the most promising candidates.

\subsection{NAS Report on Exoplanet Strategy}

The NAS Report on Exoplanet Strategy\footnote{https://www.nap.edu/catalog/25187/exoplanet-science-strategy}, released in 2018, was a consensus study by a series of committees, coordinated by the NAS, to review the state of the exoplanet field and recommend strategies for achieving future scientific goals. Similar to the ExoPTF report, the NAS Exoplanet Strategy report prominently discusses the topic of habitability. As an example of its importance, this report lists ``[learning] enough about the properties of exoplanets to identify potentially habitable environments" as an overarching goal of exoplanet science.

In its subsection \textit{What Makes a Planet Habitable?} of Section 2, the NAS Exoplanet Strategy report specifically defines its focus on ``only global or large-scale surface habitability that is potentially accessible to telescopic remote sensing." The report references the usefulness of the habitable zone, defined therein as the ``region around a star where a planet with an Earth-like atmosphere can maintain liquid water on its surface", as a useful assessment of potential habitability. But, this report consistently states that a more complete understanding of the properties of a planet and its astrophysical environment are ultimately necessary to say whether a planet can be habitable or not. Its subsection \textit{The Search for Life} in Section 3 states ``The exoplanet science community needs to expand its knowledge of intrinsic planetary properties that make habitability more likely and identify observational discriminants for those properties''. These properties include, but are not limited to, a planet's mass, geological processes, and host star activity (Figure 3.4 of the NAS Exoplanet Strategy Report).

Similar to the ExoPTF report, the NAS Exoplanet Strategy report connects habitability to the presence of surface water, but acknowledges that additional factors are also important. It recommends the development of a robust, interdisciplinary framework for assessing potential habitability, and suggests a general series of observational tests including measuring the planet's atmospheric composition, and searching for evidence of a surface ocean (e.g., via ocean glint, \citealt[][]{Williams2008,Robinson2010}).

\subsection{NAS Report Astrobiology Strategy}

The NAS Report on Astrobiology Strategy\footnote{https://www.nap.edu/catalog/25252/an-astrobiology-strategy-for-the-search-for-life-in-the-universe}, released in 2019, was a NAS-commissioned study built upon community input to provide a strategy for astrobiology research, including the search for habitable exoplanets. This 2019 report was specifically tasked with expanding upon a previous Astrobiology Strategy report led by NASA in 2015.

Unlike the ExoPTF and NAS Exoplanet Strategy reports, the NAS Astrobiology Strategy report provides a specific definition for habitability: ``an environment's ability, or inability, to support life". The report ties this ability to an environment having factors including, but not necessarily limited to, those known to be necessary for terrestrial life -- liquid water, a source of energy, and the availability of the so-called ``CHNOPS" elements. This definition is a more generalized version of the common connection between habitability and the presence of liquid water, as the latter is commonly thought to be a minimum prerequisite for supporting life. Applying this definition of habitability, then, in a quantitative assessment requires a prior definition of life, or at least what such life requires to stay alive. To this end, the Astrobiology Strategy report adopts a self-proclaimed ``life as we know it" approach. They identify life with a general set of criteria thought to be fundamental characteristics of life, including ``a means to sustain thermodynamic disequilibrium", ``an environment capable of maintaining covalent bonds ...", ``a liquid environment", and "a self-replicating molecular system that can support Darwinian evolution" (see their box 2.1, page 18).

\subsection{The LUVOIR Final Report}

The Large Ultraviolet/Optical Infrared Surveyor (LUVOIR, \citealt[][]{TheLUVOIRTeam2019}) was an ambitious space observatory concept proposed to the Astronomy \& Astrophysics 2020 Decadal Survey ``Pathways to Discovery''.
LUVOIR is designed to be a flagship mission applicable to a wide range of astrophysical domains, but particular attention to exoplanet science was paid. Finding ``habitable planet candidates'', ``confirming habitability'', and ``search[ing] for habitable worlds in the solar system'' are each prominently listed among LUVOIR's signature science cases.
Here, we review the terminology within the LUVOIR Mission Concept Final Report\footnote{\href{https://asd.gsfc.nasa.gov/luvoir/reports/}{https://asd.gsfc.nasa.gov/luvoir/reports/}}.

The report's Section 3.2.1 ``Defining habitable planet candidates" provides a definition of habitability as ``planet/exoEarth candidates [...] where liquid surface water is \textit{possible}.'' This use of habitability is subtly different than others, which actually require the presence of liquid surface water. This section further links their definition to a planet being in the habitable zone, and having an Earth-like atmosphere in terms of temperature, pressure, and composition composed of a background of N$_2$ and greenhouse gases like H$_2$O and CO$_2$. Similar to the NAS Astrobiology Strategy report, this LUVOIR report adopts a ``life as we know it" approach in defining a habitable planet as one that is confirmed to have Earth-like properties.

Their section 3.3.4 addresses ``confirmation of habitability", and what is necessary to distinguish a habitable planet candidate from a truly habitable planet. These distinguishing properties form a refined definition of a habitable planet, at least as used in that report. These include confirmed surface water, which can be achieved by observing ocean glint, for example, as well as a confirmed terrestrial mass measurement, detection of greenhouse gases including H$_2$O, CH$_4$, or CO$_2$, and signs of geological activity such as detections of volcanic sulfur-bearing species in the atmosphere. Similar to the NAS Astrobiology Strategy report, this LUVOIR report adopts a ``life as we know it" approach in defining a habitable planet as one that is confirmed to have Earth-like properties.

The LUVOIR report outlines a two-phase search and characterization strategy, using direct imaging observations, to establish a sample of habitable planets (Section 3.4). The first phase explores whether ``habitable conditions are possible on planets orbiting nearby stars'' to establish a sample of planets which are ``good candidates for habitability". The second phase simply continues this process of elimination with more in-depth observation of these select planets, including searching for biosignatures and evaluating the planet's UV-irradiation environment. In an interesting shift of language, the report refers to phase two as the ``LUVOIR habitable planet characterization program", implying that planets that clear the first phase are habitable planets. However, the only requirements that the report lists for a planet to pass the first phase, and thus be considered habitable, are being in the habitable zone, and having water vapor in the atmosphere.

\subsection{The HabEx Final Report}

The Habitable Exoplanet Observatory (HabEx) was another space observatory concept proposed to the Astronomy \& Astrophysics 2020 Decadal Survey.
Similar to LUVOIR, HabEx is designed to probe a wide realm of astrophysics, but the observatory's top science goal is to ``seek out nearby worlds and explore their habitability" (Table ES-1, \citealt{Gaudi2019}). Here, we review the terminology used in the HabEx Mission Concept Final Report.

The HabEx report avoids the use of the term ``habitable planet". Instead, the report often just refers to ``potentially habitable planet(s)'', but does not provide a definition of what makes a planet ``habitable". Based on several statements made throughout the report, including the mission's Science Traceability Matrix (STM), ``habitability" is characterized therein by the presence of water vapor in the planet's atmosphere. For example, objective 2 of the HabEx STM is ``[t]o determine if planets identified in Objective 1 (rocky planets orbiting within the habitable zone of sunlike stars) have potentially habitable conditions
(an atmosphere containing water vapor)''. This report thus seems to identify atmospheric water vapor as the primary measure of a planet's habitability.

Section 3.1 of the HabEx report outlines a general strategy for characterizing exo-Earth Candidates (defined therein as ``Earth-sized planets in Earth-like orbits") via direct imaging observations. The strategy focuses on determining the planet's orbit and searching for atmospheric water vapor and biosignatures, such as O$_2$ and O$_3$. Similar to the LUVOIR report, the HabEx report does not specifically state what observational tests a planet must pass to be considered habitable, though the report does mention a ``Follow the water'' approach. In particular, Section 3.1.2 implies that the main observational criterion is the detection of atmospheric water vapor, since the presence of water in the atmosphere is directly tied to the presence of water on the surface.

\subsection{Astro2020 Decadal Survey Report}
\label{S:Astro2020}

The ``Pathways to Discovery in Astronomy and Astrophysics for the 2020s" report \citep[][]{Astro2020}, also known as the Astro2020 Decadal Survey report, is the latest community led survey that is held every ten years to examine the state of research, identify key outstanding science questions, and outline a path for answering these questions over the next decade. This survey is sponsored by NASA, the NSF, the DOE, and the Air Force Office of Space Research to make recommendations for how the broad field of astrophysics should proceed, and has a unique influence on the direction of research funding and new mission and observatory designs.   The Astro2020 Decadal survey identified the search for habitable planets as the key theme for astrophysics in the next 10 years.

The Decadal report mentions ``habitable planet'', ``potentially habitable planet'', ``habitable planet candidate'', and similar verbiages over 150 times throughout. Despite this prevalence, the main body of the report (Sections 1 -- 7) does not specifically define the term ``habitable", nor what criteria must be met for a planet to be deemed habitable. Rather, the report seems to adopt a \textit{follow the water} approach, linking the habitability of an environment solely to that environment's ability to support liquid water. For exoplanets in particular, the ability to support liquid water is proxied by the planet's presence in the habitable zone. For example, a passage on page 76 states ``a key question [is...] the frequency of potentially habitable planets - the average number of Earth-sized planets within the habitable
zone of their star, particularly around Sun-like stars''. Similarly, the work displayed in their Figure 7.6 assumes that the frequency of potentially habitable planets is just the frequency of rocky planets within their star's habitable zone, when estimating the detection yield of potentially habitable exoplanets for future mission concepts.

The Decadal survey is assembled from reports by individual committees specializing in particular fields of astrophysics. Appendix E of the Decadal report is the Report of the Panel of Exoplanets, Astrobiology, and the Solar System. This Panel report does not further define habitability beyond what is listed in the main body of the decadal report. Relatedly, though, this Panel does stress that ``a comprehensive, systems-level approach to habitability assessment is now needed''.

One of the primary recommendations of this decade's report is a large ($\sim$ 6 m aperture) infrared/optical/ultraviolet (IR/O/UV) space telescope. Section 7 provides a basic mission concept: ``After a successful mission and technology maturation program, NASA should embark on a program to realize a mission to search for biosignatures from a robust number of about $\sim$25 habitable zone planets and to be a transformative facility for general astrophysics".

\subsection{ESA Voyage 2050 Senior Committee Report}
\label{S:ESAVoyage2050}

The 2021 report of the Senior Committee to the ESA Director of Science\footnote{\url{https://www.cosmos.esa.int/web/voyage-2050}} was tasked to lead the definition of a new long-term plan, setting out the European priorities in space science for the next couple of decades folowing the end of the previous ESA planning cycle 'Cosmic Vision'. The three main aspects of this report concern recommendations on themes that shall be addressed by the anticipated three ESA Large (L-class) missions in the coming decades, listing possible themes for Medium (M-class) missions, and long-term technology development strategies. In order to facilitate an open consultation process with the space science community, a call for White Papers and Topical Teams was facilitated to participate in the process and inform the recommendations of the Senior Committee. A community workshop with presentations from the submitted White Papers, and final reports and synthesis presentations by the Topical Teams to the Senior Committee led to a consensus on the selection of Large mission themes: (i) Moons of the Giant Planets, (ii) From Temperate Exoplanets to the Milky Way, and (iii) New Physical Probes of the Early Universe.

The Voyage 2050 Senior Committee Report mentions ``habitability'', ``habitat'', ``habitable environment'', ``habitability potential'' or similar phrases 22 times throughout. In passing, the report defines ``prerequisites for habitability'' as ``liquid water, energy, complex chemistry, and relative stability''. Within the theme (ii) recommendation for Large mission themes, the report identifies ``Characterisation of Temperate Exoplanets'' as the scientific theme with the highest scientific priority. In further descriptive details of this theme, the report specifies ``the question of the existence and distribution of life elsewhere'' and ``characterisation of terrestrial planets and their atmospheres'' as its main tasks. To infer exoplanet habitability, it recommends a tiered approach (p. 12/13): measuring radius, mass, and temperature structure of the atmosphere to decide on the rocky nature of a planet and its ability to host liquid water at its surface. Additional information on key atmospheric molecules shall deliver decisive information to test ``our understanding of the necessary conditions for a planet to develop a habitable climate''. The report further stresses the need to observe a sample of temperate planets with varying size and insolation for giving context to the measurement of peculiar abundances of some chemical species that are difficult to explain by abiotic processes.

One of the primary recommendations from the ESA Voyage 2050 report is launching a Large mission that enables the characterisation of the atmosphere of temperate exoplanets in the mid-infrared, which would enable precise measurements of radius, temperature structure, and key atmospheric molecules. In addition, the report outlines a potential Medium contribution to a large-size space telescope, such as LUVOIR or HabEx, as proposed to the Astronomy
\& Astrophysics 2020 Decadal Survey.

\section{Survey of Relevant Terminology and Frameworks}
\label{S:SurveyTerminology}

\subsection{A Powerful, Asymmetric Tool: The Process of Elimination}
\label{S:Elimination}

At the core of the deductive model of the scientific method \citep[see, e.g.,][]{popper2013realism} lies the understanding that conjectures, hypotheses, models, or fundamental theories \textit{cannot be proven right}: Instead, evidence is built up through a multitude of diverse experiments that \textit{do not disprove} them. On the contrary, a single insightful experiment can disprove even the most universal theory. This framework is relevant when we formulate hypotheses about the suitability of a planet or its sub-unit as a habitat for life, certain organisms, or ecosystems.

 In many applications of a quantitative habitability framework, a positive proof is, in fact, not required for applicability. Although not explicitly stated, all key strategic reports on the search for biosignatures in exoplanets (see Sections \ref{S:NSFExoplanetTaskForce}--\ref{S:Astro2020}) propose the \textit{process of elimination} to narrow down an initially large target sample of exoplanets to those that are likely habitable. For example, both the LUVOIR and HabEx STDT Reports describe a stepwise process in which a sequence of measurements is used to eliminate planets that are less likely to be habitable (see their Figures 1--5 in \citealt[][]{TheLUVOIRTeam2019} and 3.1--1 in \citealt[][]{Gaudi2020}).

 Although no quantitative framework is provided by these reports, the essence of the argument can be expressed as follows: The ``habitability'' of the planets depend on satisfying an unknown number of criteria, a subset of which is known (e.g., ``generally agreed upon'' requirements for life). The habitability of a planet is then expressed as:

 \begin{equation}
     P(\mathrm{Habitability} | \mathrm{Data}) = P(\mathrm{Criteria\, 1} | \mathrm{Data}) \times P(\mathrm{Criteria\, 2} | \mathrm{Data}) \times ... \times P(\mathrm{Criteria \,\, n} | \mathrm{Data}) \\
 \end{equation}

Although none of the reports provides a complete list of relevant criteria, the examples given identify the following as a sub-set of these: (1) Planetary orbit within the habitable zone; (2) Planet size similar to Earth; (3) Water vapor in the planetary atmosphere \citep[][]{Gaudi2020,TheLUVOIRTeam2019}.

Crucially, the reports all propose to use the \textit{the process of elimination} to iteratively rank the targets by their $P(\mathrm{Habitability})$ in each step of the proposed survey, and by iteratively removing those with low $P(\mathrm{Habitability})$.
Formally, this method works because in a multi-term product (where each term is between 0 and 1), \textit{the product will always be equal or lower than any given term}. Therefore, determining that even a single term has a low value necessarily places an upper limit with a low value on $P(\mathrm{Habitability})$. This process is powerful because it is applicable to \textit{arbitrary number of terms} or even to a product with unknown number of terms.

The above argument may seem trivial enough that it is rarely articulated clearly but rather assumed and used implicitly (as in the strategic reports cited above). However, the formal expression of the argument helps to clarify an essential and less trivial point: The argument is \textit{not reversible}: For a product of probabilities, \textit{no lower limit can be given} unless all terms are known. Therefore, while the process of elimination is a powerful method for even highly incomplete datasets and for highly incomplete understanding of the relevant terms, identifying planets with a high probability of being in a given state (e.g., ``habitable'') requires a complete understanding of the system studied and data of very high completeness level.

\subsection{Some Relevant Frameworks}

There is a consensus among astrobiologists that observational evidence can be used to assess the suitability of a habitat for life and that this context will be a crucial part of strategies to search for life \citep[e.g.,][]{Seager2013}. Few studies, however, attempted to synthesize considerations into an applicable, quantitative framework. In the following, we review some important examples.

\textbf{Habitability as Energy Supply vs Demand:} \citet[][]{ShockHolland2007} advocate for a habitability quantification based on energy availability and demand. They argue that the ultimate goal of life is growth, and that growth is only possible when organisms have more energy available to them than required by their power demands. A key component of an organism's power demand is the energy required to respond to stresses -- from environmental to ecological. Therefore, \citet[][]{ShockHolland2007} argue, the ultimate unit of habitability is energy/organism. They propose that studies of organisms in their environments will enable quantifying the energy balance and the growth rate as a function of the ratio of energy supply over energy demand. A similar approach -- although qualitative -- has been applied to halophiles and could serve as a template for a quantitative habitability framework.

\textbf{Relative Habitability for Transiting Exoplanets:} Soon after the discovery of the first small transiting exoplanets by NASA's Kepler mission, \citet[][]{Barnes2015} introduced a comparative method to provide relative assessment of the likelihood of the habitability of these worlds given the data available for them (typically incident radiation, orbital eccentricity, planet size). They expressed the planets' relative  habitability through a ``habitability index for transiting exoplanets'', that represents the relative probability than an exoplanet could support liquid surface water.

\textbf{Habitat Suitability:} \citet[][]{Mendez2021} proposed adopting the ``habitat suitability'' measure used in ecology as planetary habitability. In this approach, a habitat's suitability to support a specific species or group of species is described as a function. This function describes the \textit{carrying capacity} of a habitat, defined as the maximum stable biomass level it can sustain ($m_{s,max}$), where the $s$ subscript identifies the species (or ecosystem) that is modeled. The metabolism-specific function, which we will denote here $S_s$, connects the state vector of the planet ($\vec{p}$) to the carrying capacity: $m_{s,max}=S_s(\vec{p})$.
Advantages of this approach are that: (a) It allows for flexibility and complexity when describing habitat conditions for different types of life; and (b) It directly builds on terminology and methodology already in use in terrestrial ecology, thereby offering both a well-understood framework and results of decades of relevant studies on terrestrial ecosystems. {A limitation of this approach is the fact that habitat suitability indices are often difficult to construct even for terrestrial organisms: Tolerance limits to environmental stresses may be coupled \citep[e.g.,][]{Harrison2013}, leading to a complex, multi-dimensional functions that are difficult to predict based on limited data.} For our purposes, however, this approach provides a complex and often impractical-to-implement criteria as a general replacement for habitability, given that the $S_s$ functions are not understood for non-Earth-like life and may differ substantially for different types of life.

\textbf{Ecosystem Viability:} \citet[][]{Sauterey2020} modeled an ecosystem as a set of organisms performing different catabolic (energy-producing) reactions and coupled it to climate, atmospheric photochemistry, and ocean chemistry. Their model dynamically selected metabolisms and parameter values that corresponded to populations that were viable given abiotic conditions. Hence, ecosystem-scale viability was used to determine global habitability, defined by abiotic geochemical processes, as well as by the ecosystem feeding pack onto the global environment, \textit{e.g.,} by producing or consuming greenhouse gases.
Extending on this work, a formal definition of population viability \citep[being the unit of habitability][]{cockell2016habitability} was proposed as null biomass not being a globally stable equilibrium \citep{ affholder2022coupling}.

Using similar metabolism-based population modeling,  \citet[][]{Affholder2021} introduced a Bayesian framework to assess the likelihood of methanogenesis in the ocean of Enceladus.
They predicted escape rates of methane and dihydrogen in Enceladus’s plume under various hydrothermal abiotic production rates of these molecules, and under the hypothesis of an extant population of methanogens when sufficient abiotic dihydrogen was produced.
They compared these predictions with values of escape rates estimated from mass spectrometry data obtained by the Cassini mission. Their analyses showed that the Cassini observations were not in disagreement with the hypotheses of an extant biosphere at Enceladus’s seafloor, despite abundant dihydrogen (a substrate of the hypothetical methanogens) found in the plume, whereas observations were in disagreement with their abiotic model.

\section{The NExSS Quantitative Framework for Habitability}
\label{S:NewFramework}

We will now introduce a framework that enables a self-consistent, quantitative assessment of the suitability of a habitat for life (and, thus, habitability), while also satisfying the criteria identified and discussed previously. We will first clarify the terminology that the framework requires (\S\,\ref{S:Terminology}). In the following subsection (\S\,\ref{S:HabitatSuitability}), we return to and extend the description of habitat suitability. Next, in Section~\ref{S:Framework} we will introduce the framework and explain the connections between its key components.

\subsection{Terminology}
\label{S:Terminology}

We start our quantitative framework for habitability by summarizing and clarifying the terms relevant for the framework. We note that the terminology clarified here, in most cases, corresponds to the most common use of the terms in the literature, but we also acknowledge that some workers used some of these terms with somewhat different meanings. Table~\ref{T:Terms} summarizes the terms and their meaning in our proposed framework.

Our terminology includes terms that refer to the general properties of the planets in question (distinguishing \textit{Earth-sized} from \textit{Earth-like} planets), which are sometimes confused in the literature but are not interchangeable.
Perhaps the important component of our terminology relate to the \textit{types} of life that may be considered, as this implicitly carries with it the  conditions required to support it: i.e., habitability and habitat suitability are dependent on the types of life one considers. To distinguish and allow for different options, we included the definition for ``Earth-like life'' as ``life that is biochemically similar to life on Earth'' (often also referred to as ``life as we know it'').
For example, a type of life that is carbon-based, uses water as a solvent, and a quasi-periodic biopolymer for information storage and transfer could be considered Earth-like, while silicon-based life would not be considered Earth-like.

Naturally, a broader category is ``life'' itself, which would include all types of life regardless of whether or not their biochemistry is similar to Earth. As discussed thoroughly in Section~\ref{S:DefinitionOfLife}, however, there is currently no widely accepted definition of life and perhaps there never will be one. Correspondingly, our table and terminology does not attempt to define life: Rather, it introduces a set of terms that do not address habitability universally, but in a more restricted sense.

\begin{deluxetable*}{l p{6cm} p{6cm}}
\tablecaption{Summary of relevant terms \label{T:Terms}}
\tablehead{\colhead{Term} & \colhead{Definition} & \colhead{Notes} }
\startdata
{\bf Life}  & -- & No widely accepted, universal definition.  \\
{\bf Earth-like life}  & Life that is biochemically similar to life on Earth. & Equivalent to "life as we know it". \\
\hline
{\bf Rocky Planet}  & A planet consisting dominantly of refractory elements. \\
{\bf Earth-sized Planet}  & A planet with a diameter similar to that of Earth.  & May differ from Earth in its other properties. \\
{\bf Earth-like Planet}  & A planet known to resemble Earth in its properties and key processes: As a system, it can be expected to respond to changes similarly to how Earth would respond (in its corresponding era). &  An Earth-like planet is not necessarily inhabited.  \\
\hline
{\bf Habitable Zone}  & The spherical shell around a star where an Earth-like planet could have stable liquid water on its surface. & \\
{\bf Metabolisms} & A term representing organism(s) that share primary metabolic pathway(s). &  May refer to a single cell, a cluster of cells, an organism, a species, or an ecosystem.\\
{\bf Viability model} & A model that describes the necessary conditions for a metabolism to complete its life cycle and sustain a stable population. & \\
{\bf Suitable Habitat for X} & An environment with conditions that meet the necessary requirements for viability for metabolism X.\\
{\bf Habitat Suitability}  & The measure of the overlap between the necessary environmental conditions for a metabolism  and environmental conditions  in the habitat.  & Provides more customized assessment for a specific type of metabolism, but not universally applicable.  \\
\hline
\enddata
\end{deluxetable*}

\subsection{Habitat Suitability}
\label{S:HabitatSuitability}

Lacking a universally valid definition of ``life'', no universally valid conditions can be identified for ``habitability''. How can our community address and study the conditions required by non-Earth-like life -- or for a sub-set of Earth-like life, such as methanogens -- without an understanding of what life is? By  definition, non-Earth-like life utilizes a biochemistry that is different from Earth-like life and, thus, will possibly (even likely) require very different physical and chemical conditions. One may introduce adjectives to refine and distinguish different types of habitability. However, after considering such options, we found these cumbersome and too inflexible to meet the anticipated needs.

Instead, looking toward the next steps in the process of quantifying our terminology, we build on but modify a terminology rooted in terrestrial ecology. As proposed by \citet[][]{Mendez2021}, we adopt the use of ``habitat suitability for organism model X'' to describe whether the conditions present in a given environment allow (or are conducive to) of the viability of a metabolism. The assessment of habitat suitability is, in essence, the study of viability of metabolisms under a given set of environmental conditions.

Our terminology and use differs, however, from that introduced by \citet[][]{Mendez2021} (and many of the terrestrial ecology studies) in one important aspect: We do not express the habitat suitability as a ratio of biomass relative to the habitat's carrying capacity, but rather as a probability that the metabolism is viable in the habitat.
This difference is introduced to enable a more general application of the framework, including situations where only very limited information is available on potential habitats (which is often the case in extraterrestrial objects). While the conventional habitat suitability assessment needs to measure the sustainable growth rate of a species/ecosystem in a complex habitat (often in the presence of other organisms and feedbacks), and compare that to a theoretical (or empirical) maximum carrying capacity, our framework sets a simpler goal:
\emph{We aim to assess habitat suitability by evaluating the probability of the habitat's conditions meet the necessary conditions for the viability of specific metabolisms.}

We stress the importance of the metabolism viability model in our framework:
Lacking well-established experimental data for non-Earth-like life (and even for the majority of terrestrial organisms), the habitat suitability can only be applied with the help of a \textit{model} that describes viability of the organisms as a function of environmental conditions. Such models exists and are in use in ecology. They are typically based on terrestrial organisms and the parameterized understanding of their metabolic efficiency. Nevertheless, this framework can be readily extended to model non-terrestrial (yet hypothetical) metabolisms.

\subsection{The Structure of the Framework}
\label{S:Framework}

\begin{figure}[ht!]
\begin{center}
\includegraphics[width=0.94\linewidth,angle=0]{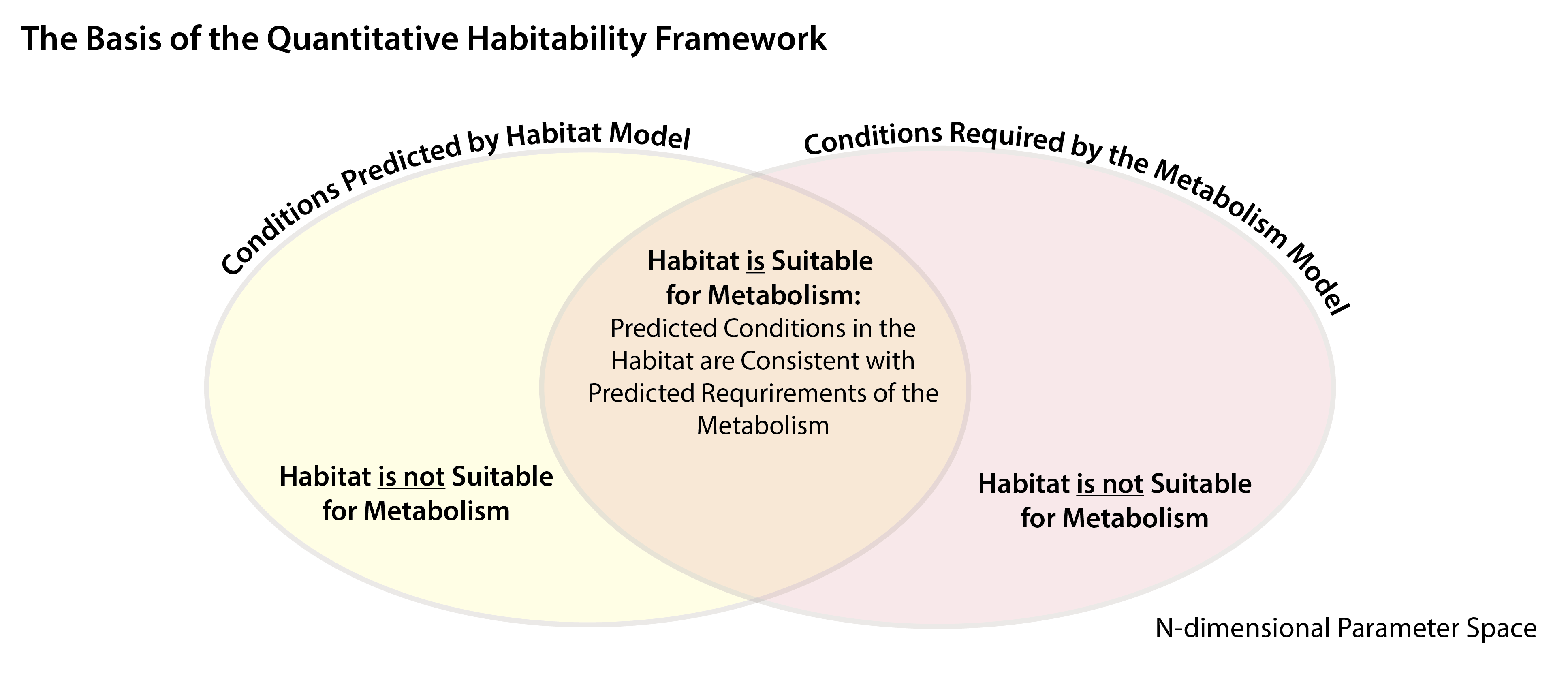}
\caption{Illustration of the basis of the Framework for Habitability: The comparison of the environmental conditions predicted by the habitat model and the environmental conditions required by the metabolism model.  \label{f:Basis}}
\end{center}
\end{figure}

\begin{figure}[ht!]
\begin{center}
\includegraphics[width=0.97\linewidth,angle=0]{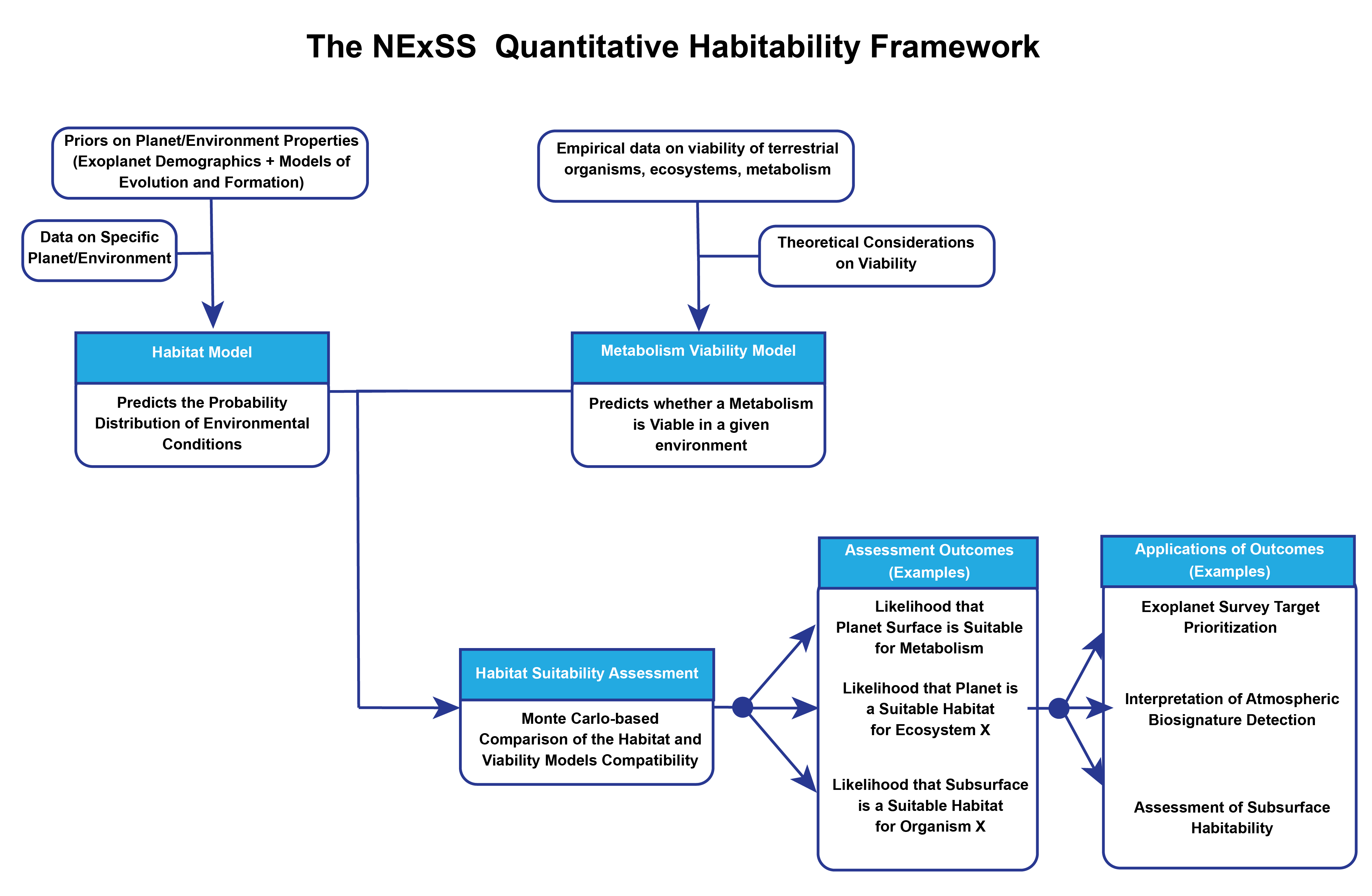}
\caption{Overview of the NExSS Framework for Habitability. The framework assesses the suitability of habitats for model organisms/ecosystems. The potential habitats are described through a combination of system/planet-specific data and priors informed by exoplanet statistics. Habitat suitability is assessed by applying a function specific to species/organisms to the habitat properties. The outcomes are probabilistic in nature. This framework is both flexible and specific. A special case of habitat suitability is included that corresponds to ``surface liquid habitability''.   \label{f:Framework}}
\end{center}
\end{figure}

In our framework we compare two models: One for the conditions in the habitat and one for the requirements of the metabolism. At the very fundamental level, the likelihood of the habitat being suitable for the metabolism is provided by this quantitative comparison (Figure~\ref{f:Basis}). Figure~\ref{f:Framework} provides a more detailed overview of the \framework{} proposed in our study. In essence, the framework is defined to manage the flow of structured information through a model that provides a quantitative assessment of the suitability of the potential habitat considered to support the entity (organism, species, ecosystem) that is modeled.

The quantitative information describing the potential habitat considered is synthesized from two sources: (A) Priors on the habitat (planet) properties, which could include distributions from population-level trends, or from model predictions, etc. (B) Data on the specific habitat (or planet) considered, which could include probability distributions for planet mass and radii, stellar irradiation or surface temperature, constraints on atmospheric composition, etc. Information (typically in the form of probability distributions) from sources A and B are combined to provide a probabilistic assessment of the habitat's (or planet's) state -- which we will refer to as the \textit{habitat's state vector} or H, which is -- considering uncertainties in the properties --  an n-dimensional probability distribution (one dimension for each of the parameters considered).
The habitat's state vector is an input to the \textit{habitat suitability assessment} module. This module is central to the framework, as it provides a probabilistic comparison of the properties of the habitat (as reflected by the habitat's state vector) and the needs of the metabolism assessed.
The metabolism could be an individual organism (e.g., a specific human), a species (e.g., \textit{Emiliana huxleyi}), or an ecosystem (e.g., the Sonoran desert) (for a discussion of this aspect, see Section~\ref{S:Metabolism}). The needs of the metabolisms assessed are expressed through a viability model, which we will refer to as a \textit{V-function}. V-functions make predictions for the viability of the metabolisms as a function of the habitat's state vector (e.g., environmental
conditions).

Given the above, the framework can be essentially captured as an n-dimensional volume integral of the product of the probability distributions of the habitat's state vector and the metabolism's viability function:

\begin{equation}
Q_{\mathit{S}} = \int_{-\infty}^{+\infty} P(V=1 | x_i) d x_i, = \int_{-\infty}^{+\infty} V(x_i) \times H(x_i) d x_i,\\
\end{equation}

where Q is the habitat suitability, i.e., the probability that the metabolism is viable given the habitat is in state $x_i$. Here, $V(x_i)$ is the viability function (the probability that the metabolism is viable given the specific state of the environment), and $H(x_i)$ is the habitat function (i.e., the probability that the environment is in the specific state described by the state vector).

The habitat suitability assessment, as well as its outcome, are probabilistic in nature. A convenient implementation is through the use of Monte Carlo sampling of the distributions representing the state vector of the habitat, and the application of the V-function to each of the sampled habitat manifestations. The resulting assessment will then reflect the likelihood that the considered habitat can support the modeled metabolisms.
As the resulting assessment is quantitative and should ideally consider the entirety of the relevant and available information (on the specific system, on its parent population, and on the modeled organism), it will constitute the best available assessment of the suitability, and one that can be applied to address scientific needs.

We stress, that the example shown in Figure~\ref{f:Framework} is not a universal implementation of the \framework, but is intended to demonstrate the organization of the framework components. Specific implementations will require customized modules.

{We note that the habitat module may model potential habitats over a variety of scales, locations, times -- and even spatial dimensions. For example, on one extreme, it could describe a microscopic pore in a rock  at a given instant in time. This would likely be well reproduced by a zero-dimensional, stationary model. On the other extreme, the habitat model could describe an entire planet, for which a two- or three-dimensional and potentially dynamic (time-evolving) model may be appropriate. The realism and limitations of the model used for habitat and metabolism must be kept in mind when interpreting the results from the QHF framework. In the examples that follow, we demonstrate the QHF framework functionality for zero and one-dimensional models, but these comparisons can be naturally extended to higher dimensions and to smaller and larger scales.}

\section{Example Models for Metabolisms}
\label{S:V-functions}

In this section, we introduce example V-functions (Viability functions) and demonstrate applications of the \qfw. These V-functions describe the compatibility of the environmental conditions and the conditions a life form (e.g., individual, species, or ecosystem) requires. They are, however, not intended to predict growth rates or expected biomass, or otherwise characterize how wide-spread the metabolisms may be in the environment modeled.
We will first introduce a V-function corresponding to the classical liquid water habitability criteria. This V-function will demonstrate how the NExSS Framework is compatible with many past uses of habitability in the specific sense of ``liquid water is stable on the planetary surface''.
We will introduce V-functions for {methanogens (archaea) and cyanobacteria}. The latter represents an example for \textit{Earth-like life capable of carrying out oxygenic photosynthesis on planetary scales}. Following the bacterial V-functions, we will introduce an V-function for eukaryotes.
In defining V-functions for the different types of terrestrial organisms, we follow the environmental parameter ranges compiled in Table 3.1 of the report by \citet[][]{BarossReport2007} and the excellent review by \citet[][]{Clarke2014}.

Note, that the V-functions introduced here are examples and represent starting points for models of these organisms' requirements, rather than complete and final models. {Considering this study as a proof-of-concept, w}e encourage future work to expand upon and refine the conditions included in these examples.

\begin{figure}[ht!]
\begin{center}
\includegraphics[width=0.95\linewidth,angle=0]{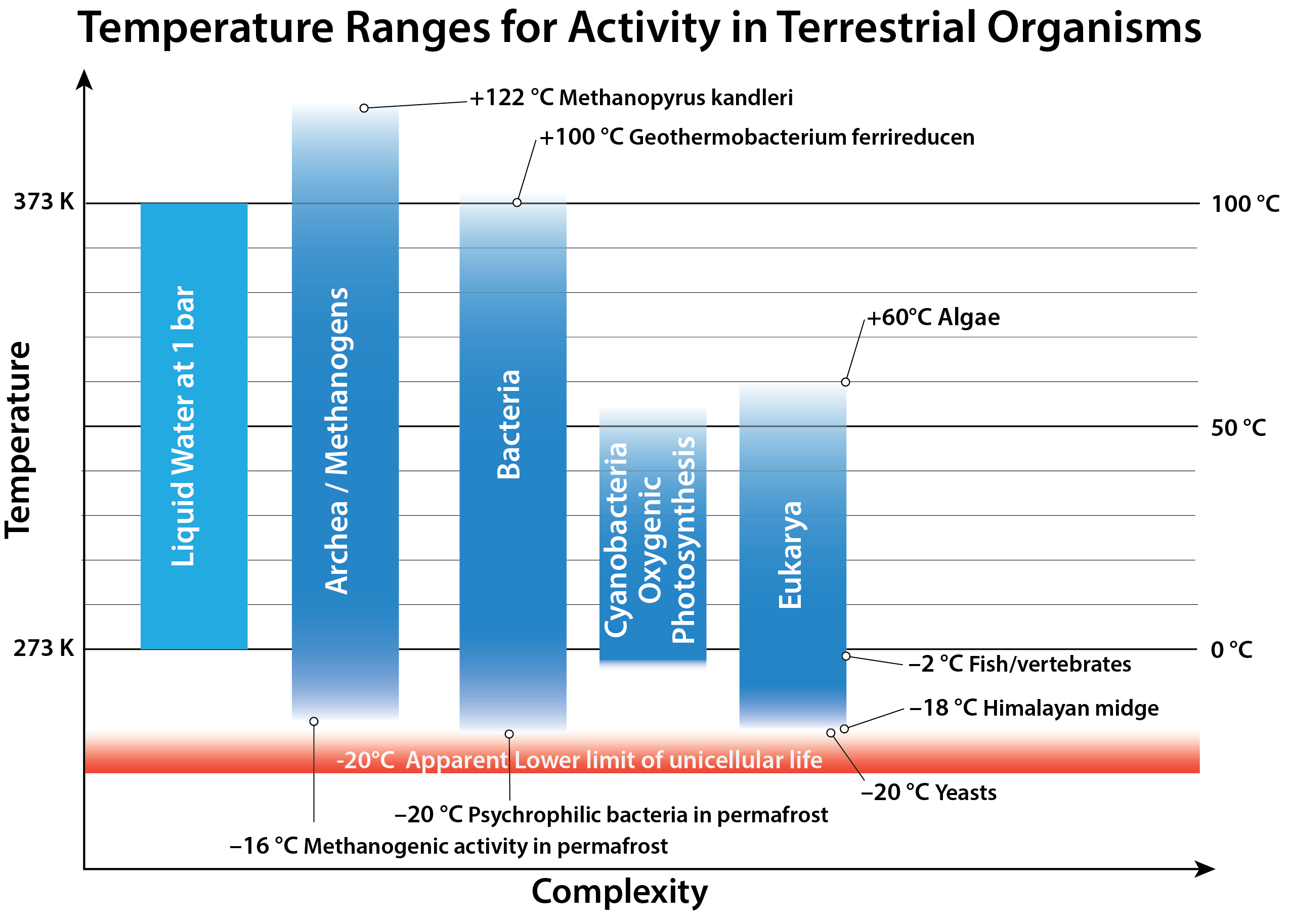}
\caption{Overview of the temperatures ranges in which representative types of terrestrial organisms are biologically active (i.e., likely to complete life cycles).  \label{f:TemperaruteRanges}}
\end{center}
\end{figure}

\subsection{Liquid Surface Water}

In the literature, perhaps the most commonly used criterion for a ``habitable planet'' is one that identifies these as ``broadly Earth-like planets with a potential to harbor stable liquid water on their surfaces''. Within the \qfw, this criterion equals a V-function that is 1 when the surface conditions are consistent with the existence of liquid water and 0 otherwise. The simplest, physically meaningful implementation of this V-function is as follows:

\begin{equation}
V_{\mathit{lH2O}} = V_{T} \times V_{\mathit{pressure}},\\
\end{equation}

\[
    V_{T} =
    \begin{cases}
        1 & \text{if 273 K $< T_{\mathit{water}} <$ 373 K,}\\
        0 & \text{otherwise.}
    \end{cases}
\]

\[
    V_{pressure} =
    \begin{cases}
        1 & \text{if 0.006 bar $< P_{\mathit{surface}}$,}\\
        0 & \text{otherwise.}
    \end{cases}
\]

In our initial model presented here, we adopted this implementation. We caution that this simplistic approximation neglects {the composition of water, which may impact its phase transition conditions (see below)}.

Indeed, liquid water stability is dependent on pressure and temperature. At the most fundamental level -- without taking into account the effect of solutes that can lower the water activity and increase the liquid stability range -- liquid water stability field is bordered by the saturated vapor pressure curve and the freezing curves of ices that both have strong pressure dependency. The saturated vapor pressure curve can be calculated following \citet[]{Buck1981} with the approximation from the triple point at P=0.00611657 bar and T=273.16 K to the boiling point of water at T=373.15~K at 1 bar pressure:

 $$P [bar] = 0.0061121 \cdot \exp\left(\left(18.678-\frac{T [^\circ C] +273.15~K}{234.5~K}\right) \times \frac{T[^\circ C]+273.15~K}{T[^\circ C]+530.29 K}\right)$$

At depth in a putative ocean the increasing pressure can also affect the stability of the liquid water. Initially, increasing pressure depresses the melting curve of ice \textsc{Ih} from 273.15 K at 1 bar pressure to 251K at 2,090 bar pressure at the ice \textsc{Ih}--\textsc{III}--liquid triple point following the following Simon–Glatzel equation:

$$P (bar) = -39.52 \left(\left(\frac{T[K]}{273.16}\right)^{9}-1\right) $$

Above that pressure Ice \textsc{III}, then \textsc{V}, then \textsc{VI} then \textsc{VII} will be present at the melting point which will have a positive Clapeyron slope. They are not considered here, as liquid water will always be present if the temperature is above the melting line of ice \textsc{Ih} below 2,090 bar pressure.

Water activity and pH are two other parameters that can significantly affect the viability of aqueous environments. Water activity, often refereed as $a_w$, is a thermodynamic measure of the amount of non-chemically bounded water (i.e., amount of water available), which decreases when an increasing amount of solute is dissolved. Water activity is a solute-agnostic metric crucial in food science to estimate the safety of preservation, and therefore can be used in the present study to bound the maximum range of viability. pH is a chemical measurement of {how acidic (pH$<$7) or  basic (pH$>$7) a solution is}. Various geochemical processes can affect the pH of aqueous environments on Earth, Mars and other planets from very acidic to very {basic}.
{Furthermore, the presence of salts can depress the freezing point of water, extending the temperature range in which it is liquid. Such localized brines are common in the ices of Antarctica, on Mars and the icy moons \citep[e.g.,][]{Martinez2013,Chivers2023}. If salt-laden water is more common than freshwater, then the lower freezing point they enable may significantly increase the volume of viable habitats in many locations. Therefore, in future studies we plan to incorporate a more comprehensive treatment of water activity in the water phase transition models.}

In the following, we describe the known and conservative viability temperature limits
for Archaea, bacteria, and cyanobacteria.

\subsection{Archaea incl. Methanogens}

Archaea are unicellular prokaryotic organisms that are often also extremophiles. Archaean metabolisms include methanogenesis, a {range of} chemotrophic pathway{s that evolved early on in Earth's history and that are} fundamentally independent of insolation (i.e., they may be relevant in habitats without light). Furthermore, methanogen{ic metabolic pathways} require less complex biochemical processes than oxygenic photosynthesis. Therefore, {early metabolisms such as methanogenic ones} may be more common in the universe than oxygenic photosynthesis. Methanogenic {microbes utilize diverse electron donors and carbon sources, but all produce methane as a byproduct of their metabolism. Production of methane by these organisms is of } interest as they have the potential to produce it on planetary scales {which could be} detectable \citep[e.g.,][]{Sauterey2020,Thompson2022}

Our high-temperature limits for archaea are informed by hyperthermophilic archaea common in hydrothermal vents \citep[e.g.,][]{stetter2007hyperthermophilic}. \textit{Geogamma barossi} strain 121 \citep[][]{KashefiLovley2003} -- isolated from the Mothra hydrothermal vent field in the Northeast Pacific -- was shown to complete its lifecycle at 121\celsius. Similarly, a strain of \textit{Methanopyrus kandleri} -- isolated from the Kairei vent field on the Central Indian Ridge -- was found to grow at temperatures up to 122\celsius{} \citep[][]{Takai2008}.
The low-temperature limit for archaea is informed by the discovery of methanogenesis in permafrost cores \citep[][]{Rivkina2007}, likely linked to the archaea \textit{Methanosarcina mazei}. Methanogenesis was measured to continue down to $-$16.5\celsius{}.

\[
    V_{archaea}(T) =
    \begin{cases}
        1 & \text{if 257 K $< T <$ 395 K,}\\
        0 & \text{otherwise.}
    \end{cases}
\]

{These limits approximate the range of temperature conditions in which archaea that employ various metabolic pathways are viable, including but not limited to methanogenic ones (G. barossi is an iron reducer; M. kandleri is a hydrogenotrophic methanogen; M. mazei can either utilize {H$_2$}/CO$_2$ or acetate). While specific metabolisms may not span the entire temperature range used here, methane-producing archaea are known to exist throughout this range.}

{In addition to the temperature, methanogen viability may also be limited by hydrostatic pressure. As of now, there is no well established ultimate pressure limit for life but, in terrestrial environments, oceanic vents provide environments with high temperatures and high pressures \citep[]{Yang2021}{}. These environments informed our choice of upper pressure limit. For the purposes of consistency, we adopt here an upper pressure limit of 55~MPa. \textit{M. kandleri} has been shown to grow at this pressure, with optimal growth occurring at 20~MPa \citep[][]{Takai2008}. 55~MPa is the highest known {natural} pressure tolerance of methanogens, which have only been isolated down to depths around 2,000 meters, similar to \textit{M. kandleri}.
{Laboratory experiments on the methanogen \textit{Methanococcus jannaschii} have shown that, under specific temperature and chemical conditions, growth may still occur up to pressures of approximately 75~MPa \citep{Miller1988}, but we reserve this value for speculative examples.}
In contrast, the most piezophilic organism currently known is \textit{Colwellia marinimaniae}, a bacterium isolated from the gut of an amphipod in Challenger Deep \citep[][]{Kusube2017}. \textit{C. marinimaniae} has shown growth at a maximum pressure of 140 MPa, with optimal growth at 120 MPa. This discrepancy between hyper-piezophilic species and piezophilic methanogens may potentially be attributed to a sampling bias, or some physical or environmental limitation of the methanogenic metabolism. Given the above, we conservatively adopt 55~MPa as upper pressure limit for methanogens:}
\[
    V_{archaea}(P) =
    \begin{cases}
        1 & \text{if P $<$ 55~MPa,}\\
        0 & \text{otherwise.}
    \end{cases}
\]

\subsection{Bacteria}

Some bacteria can grow at very high temperatures, although no known bacteria can match the high-temperature tolerance of hyperthermophilic archaea. The upper temperature limit for our bacteria model is set by \textit{Geothermobacterium ferrireducens}, which was isolated from the Obsidian Pool of Yellowstone National Park and grows at 100\celsius{} \citep[][]{Kashefi2002}. The lower temperature limit for bacteria is less well established, but \citep[][]{Rivkina2000} found evidence for psychrophilic bacterial respiration in permafrost cores at temperatures as low as $-20$\celsius. We adopt this temperature as the lower limit for bacterial life.

\[
    V_{bacteria}(T) =
    \begin{cases}
        1 & \text{if 253 K $< T <$ 373 K,}\\
        0 & \text{otherwise.}
    \end{cases}
\]

\subsection{Cyanobacteria}
\label{S:Cyanobacteria}

Next, we focus on cyanobacteria, a particularly important phylum of bacteria, and one with a significantly narrower set of environmental parameters for viability than the generic bacterial V-function. Cyanobacteria played (and continue to play) important roles in the coevolution of life and our planet, and similar organisms may perhaps shape other inhabited planets, too. Stromatolites, fossilized, macroscopic, sediment-rich, layered bacterial mats, date back to about 3.5~billion years ago and provide the earliest indisputable evidence for life \citep{Walter1980}. {However, uncertainties remain in the details of the sedimentation process and how {Archean} stromatolites} compare to potential modern-day analogs \citep[e.g.,][]{Bosak2013}. Nevertheless, it is widely accepted that the bacterial mats were formed by cyanobacteria. This assumption is important as, unlike other photosynthetic bacteria, modern cyanobacteria utilize water as an electron donor and -- using Photosystems {\sc{I}} and {\sc{II}} -- release oxygen during photosynthesis \citep[e.g.,][]{Sanchez-Baracaldo2020}.

About a billion years after the formation of the earliest known stromatolites, Earth's atmospheric composition changed with the accumulation of free O$_2$ \citep{Farquhar2000,Lyons2014}.
It is likely that cyanobacteria has driven or significantly contributed to this change \citep[e.g.,][]{Cloud1965,Fischer1965}, also other factors, such as geologic evolution of the Earth's crust, may have played important roles \citep[][]{Gaillard2011,Lyons2014,2017NatGe..10..788S}.
This Great Oxidation Event (GOE) changed not only the composition of Earth's atmosphere but also the composition of its lithosphere \citep[e.g.,][]{Sverjensky2010}. To this day, cyanobacteria remain major contributors to the oxygen production in the terrestrial ecosystem. Importantly, cyanobacteria also contributed to the emergence of algae and plants: All algae and plants use chloroplasts for photosynthesis. Chloroplasts originated from endosymbiotic cyanobacteria and, to this day, contain their own genomes and genetic systems \citep[][]{Allen_ChloroplastGenome_2015}.

We define here an initial V-function for the phylum cyanobacteria, which we identify as version AE v1.0. Our V-function aims to capture the ranges of the combined parameters within which cyanobacteria are viable.
In this initial model, we consider the following key parameters: Temperature range and pH levels.
For the temperature range for cyanobacteria, we adopt 0\celsius{} to 70\celsius, following Figure 3 in \citet[][]{Clarke2014} (who used data from \citealt[][]{brock2012thermophilic}, originally published in 1978). This upper limit is based on data showing that species richness in Yellowstone geothermal pools for cyanobacteria declines strongly for T$>$50\celsius{} and reaches near-zero values around 70\celsius.

\begin{center}
\begin{equation}\label{Eq:Cyano}
V_{cyano}= V_T
\end{equation}
\end{center}

\begin{equation}\label{Eq:CyanoConditions}
    V_T =
    \begin{cases}
        1 & \text{if 273 K $< T <$ 343 K,}\\
        0 & \text{otherwise.}
    \end{cases}
\end{equation}

\subsection{Eukarya including Plants and Fungi}

Eukarya include organisms with a broad range of complexity, ranging from unicellular to complex, multi-cellular plants, fungii, and animals. In our discussion, we exclude the special case of endotherms, i.e., organisms that actively stabilize their internal temperature (e.g., mammals and birds).

The low temperature limits of non-endothermic eukarya are similar to those of archaea and bacteria:
Most eukarya are limited to temperatures above water's freezing point ($>0$\celsius), but some examples of activity at temperatures approaching $-20$\celsius{} have been reported \citep[][]{Clarke2014}. Nevertheless, the universal lack of spoilage of foods refrigerated to $-20$\celsius{} and below suggests that no eukarya survive at those temperatures.

We note that, the sturdiest plants are likely lichens (symbiotic relationship between fungi and algae). Lichens tolerate {low temperatures and wind desiccation} and were shown to photosynthesize at temperatures as low as $-16$\celsius, and even under Martian conditions \citep[][]{deVera2008,Onofri2012}.
On the high-temperature end, eukaryotes are less resistant than bacteria and archaea.

\section{Example Applications}
\label{S:Applications}

In the following, we review four examples that demonstrate the broad range of applicability of the \framework{}. These examples are general enough to apply to different types of observations and instruments, but specific enough that their use context and derived requirements can be established with reasonable detail. For exoplanets, we will study a case of exoplanet target prioritization for follow-up (Example 1), and the interpretation of biosignature observations in the context of habitability (Example 2). To demonstrate Solar System use cases, in Example 3 we explore the habitability of the Martian surface and sub-surface, while in Example 4 we demonstrate the framework on the habitability of Europa's ocean. Table~\ref{T:ExampleResults} provides a high-level summary of the configuration of the \framework{} for each example and the result of the habitat suitability assessment.

\subsection{Example 1: Target Prioritization: a \TRe{} vs \TRf{} Planet?}
\label{S:Example1}

This case exemplifies situations when, as part of a dedicated exoplanet search-and-characterization mission, the potential target list must be narrowed down, i.e., targets must be prioritized to enhance the likelihood of mission success. The prioritization may occur as part of the survey definition, while re-prioritization may be desirable during the mission. The quantitative assessments of the potential habitability of planets in the targeted systems are among the key considerations for the target prioritization. Depending on the specific mission goals, the assessment of habitability may be combined with other factors (such as observability of targets or the availability of ancillary information). For examples of this approach, see \citet[][]{Seager2013,Gaudi2019,TheLUVOIRTeam2019,BixelApai2020}.

In Example 1, we compare two planets of interest with incomplete information and aim to prioritize them for in-depth spectroscopy. The goal is to focus on the planet that is more likely to provide suitable conditions for methanogens. We assumed here that basic data are available on the host star (luminosity, radius), and on the planets (orbital parameters), but that the atmospheres' compositions remain unknown.
Lacking information on the planets' atmospheres, we assumed the same planetary atmosphere model for both.
{To provide relevant examples, we base our models on TRAPPIST-1e and TRAPPIST-1f but do not aim to build models that provide the best representions of those specific planets.}
We used the following properties for the host star TRAPPIST-1: {stellar mass} $M_{*}=0.0898\pm0.0023 \, M_\odot$ \citep[][]{Mann2019} and stellar luminosity $L_{*}=0.000553  \pm 1.92\times10^{-5}\, L_\odot$ \citep[][]{Agol2021}. We used the following parameters for our \TRe{} planet model:
{orbital semi-major axis} $a=0.02925\pm0.00025$ au, {orbital period} P=$6.101 013\pm0.000035$\,d. The following parameters were adopted for our \TRf{} planet model:
$a=0.03849\pm0.00033\,$au, and P=$9.207 540\pm0.000032$\,d. These planet parameters are from \citet[][]{Agol2021}.

For illustrative purposes, we will use a simple energy balance model for both planets:
We calculate the equilibrium temperature of the planet assuming an isothermal {planet with a Bond albedo ($A_B$)}, and a simple greenhouse model (see below). The equilibrium temperature of the planet is calculated as:
\begin{equation}
    T_{eq} = \left( \frac{L_* (1-A_B)}{16 \pi a^2 \sigma} \right)^{1/4}
\end{equation}

 where $A_B$ is the Bond albedo, {$a$ is the orbital semi-major axis}, $\sigma$ is the Stefan-Boltzmann constant.
 {(Our model assumes that the planetary surface is a blackbody emitter.)}
A single-layer leaky greenhouse model, in which parameter $\alpha_L$
describes the fraction of the thermal radiation emitted by the planetary surface that is absorbed and re-emitted by the atmosphere \citep[e.g.,][]{Seager2010exoplanet}. Then, the surface temperature $T_s$ is higher than the radiative equilibrium temperature $T_{eq}$:

$$T_s = \left( \frac{2}{2-\alpha_L}  \right)^{1/4} T_{eq}$$

{This simple model is a good approximation for thin (Earth-like) atmospheres orbiting broadly Sun-like stars. For example, a case with $\alpha=0.0$ would correspond to no greenhouse warming, while an $\alpha$=1.0 case would correspond to the maximum greenhouse warming in this model ($T_{s}=1.189\times T_{eq}$). This model reproduces Earth's surface temperature and amplitude of greenhouse warming for an $\alpha=0.77$ value. However, this simple model has multiple caveats, especially when applied to planets markedly different from Earth. For example, Venus-like (thick CO$_2$-rich) atmospheres will introduce greenhouse warming that far exceeds what the above single-layer atmospheric model can reproduce. In such cases, the surface temperature will be underestimated by the model. Another simplification in the model is the assumption that the atmosphere does not absorb stellar radiation. While a good approximation for Earth, this assumption is less valid for cool stars where stellar emission peaks closer to potential atmospheric molecular absorption bands, leading to enhanced atmospheric absorption \citep[][]{Shields2013}.}

{While future studies can introduce more complex atmospheric models to address the above caveats and make more reliable predictions for specific planets, our goals here are to illustrate the \framework{} and thus adopt this model while noting its caveats.}
As the properties of the planets in question are yet unexplored, we assume a normal distribution for $\alpha$ with a mean value of 0.77 and 1$\sigma$=0.05, intended to represent a present-day Earth-like atmosphere that is not necessarily identical to Earth. We limit $\alpha$ to be between 0.0 and 1.0.
{The albedo of the planet is an important parameter that impacts the equilibrium and surface temperatures. The albedo is set by a combination of the stellar spectrum and the wavelength-dependent opacity (gas-phase and particulate) of the planetary atmosphere and surface. Even for a broadly Earth-like planet, the albedo may vary with the stellar temperature and  wavelength of peak instellation \citep[e.g.,][]{Shields2013}. As discussed in that study, energy balance and general circulation models predict a broad range of possible albedos (from 0.2--0.8).
Similarly, Solar System planets with atmospheres and Titan, as well as warm/hot sub-neptune exoplanets  span a broad range of albedos \citep[][]{Basant2022}. To reflect this range of possible albedos, we adopt a normal distribution as prior for the $A_B$ Bond albedo with mean $A_B=0.3\pm0.1$ (very similar to a Bond albedo distribution derived from exoplanet observations by \citealt[][]{SheetsDeming2014}).}

\begin{figure}[ht!]
\centering
\includegraphics[width=0.75\textwidth]{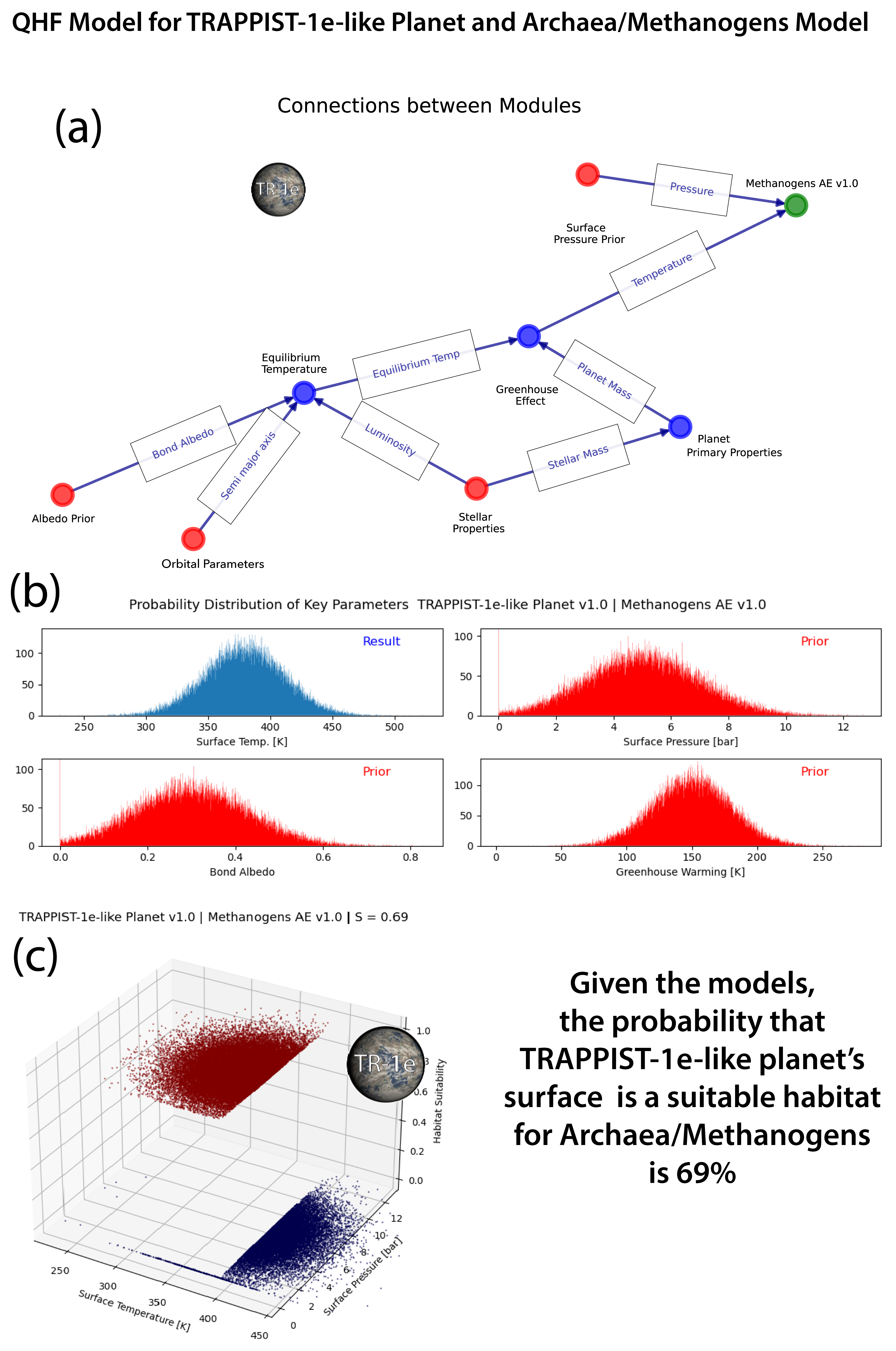}
\caption{QHF assessment of the viability of Archaea/Methanogens in a modeled \TRe{} planet's surface habitat. {a:} Connections between the model modules. Red are priors, blue are calculated values, green is the viability model. {b:} Relative probability distributions of key parameters. {c:} The distribution of calculated viability as function of surface temperature and pressure. \label{f:Example1_TRAPPIST1e}}
\end{figure}

\begin{figure}[ht!]
\centering
\includegraphics[width=0.75\textwidth]{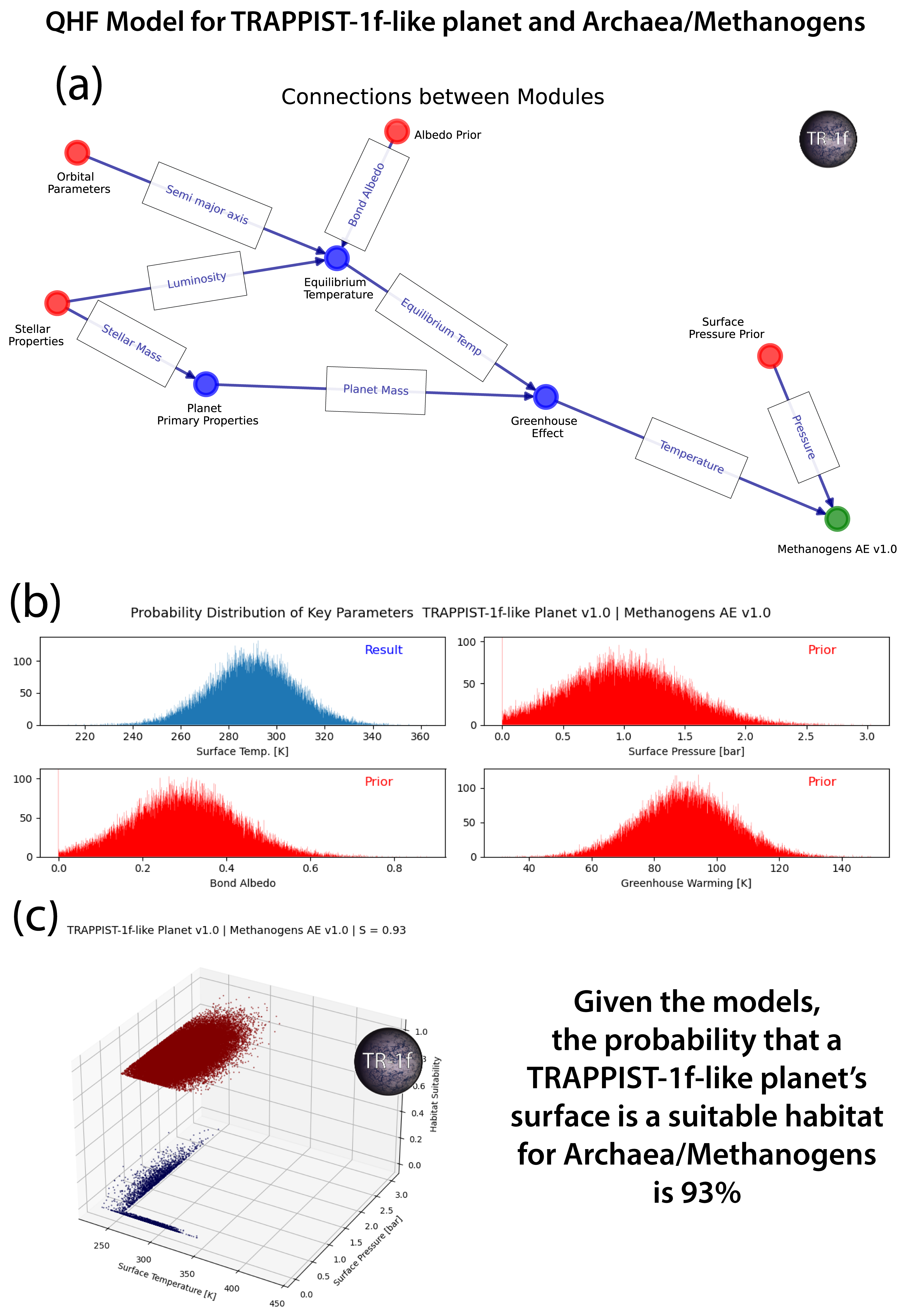}
\caption{Same as Figure~\ref{f:Example1_TRAPPIST1e} but for a \TRf{} planet. \label{f:Example1_TRAPPIST1f}}
\end{figure}

Figure~\ref{f:Example1_TRAPPIST1e} provides an overview of the connections between the modules that define the habitat state and the metabolism's needs. Red nodes are priors, blue nodes are calculated. The arrows indicate the direction of information propagation. In the final (green) node (Methanogens AE v1.0), the habitat's state is compared to the metabolism's needs to calculate the probability of the metabolism's viability in the environment, i.e., the habitat suitability.

In this model, the priors (Bond albedo, orbital parameters, stellar parameters) determine the radiative equilibrium temperature. The {planet's atmospheric composition} influences the greenhouse effect which -- in combination with the radiative equilibrium temperature -- sets the surface temperature of the planet. {The surface pressure is a prior. Currently, the atmospheres of TRAPPIST-1e and -1f are not well understood. For our demonstrations, we set the pressure priors (within realistic boundaries) to illustrate interesting model behaviors. Specifically, we assume normal probability distributions for the pressure. For \TRe{} planet model we set the pressure to be 5$\pm$2 atm$>$0, and for the \TRf{} planet model we set it to be  1$\pm$0.5 atm$>$0. These ranges encompass Earth-like pressure levels that are consistent with the simplified energy balance equation used above, while allowing for differences between the planets.} The surface temperature and surface pressure are compared then to the metabolism model (Methanogens AE v1.0) to determine the viability of the metabolism on the surface of the planet. In the implementation included with this manuscript, the graph is built automatically via an algorithm. The key parameters are represented by probability distributions (i.e., uncertainties) and a Monte Carlo-based assessment is used to propagate the probability through the graph nodes.

{We emphasize that the goal of our model is to illustrate the \framework{} and do not aim for completeness: We specifically do not consider climate feedbacks, such as the carbon-silicate negative feedback that stabilizes Earth's climate through the regulation of the partial pressure of CO$_2$. Consequently, our model does not aim and it is not expected to reproduce the boundaries of the habitable zone as models with atmospheric chemistry and climate feedback cycles would \citep[e.g.,][]{Kopparapu2013}. Nevertheless, our \framework{} is compatible with such models and future applications could include such integration.}

As an example, Figure~\ref{f:Example1_TRAPPIST1e}b shows the distribution of two priors (Bond albedo and surface pressure), as well as two calculated values (greenhouse warming and surface temperature) for {a \TRe planet}. Figure~\ref{f:Example1_TRAPPIST1e}c shows the overall habitat suitability assessment for {a \TRe planet}. The figure shows habitat suitability (0 vs 1) as a function of surface temperature and pressure evaluated in Monte Carlo realizations of possible planet states. Blue dots represent possible states that are incompatible with the viability function of our Methanogens (AE v1.0) model, while red dots represent states where the habitat and viability functions are compatible. The fraction of viable states is the probability that the metabolism is viable in the habitat.

We carried out an identical assessment for a {\TRf planet} model, using the appropriate planet parameters. The results of that assessment are shown in Figure~\ref{f:Example1_TRAPPIST1f}. Its three panels show the connections between the modules (panel a), the relative probability distributions of four key parameters (panel b), and the resulting viability assessment as a function of surface temperature and pressure.

The high-level results in our examples are as follows: Given our models and assumptions, we find that the probability that the surface conditions on a {\TRe{} planet} are consistent with those required for the viability of Methanogens is 69\%. For {\TRf{} planet model, the corresponding} probability is 93\%.

\subsection{Example 2: Interpretation of Biosignatures in Context of Habitability}
\label{S:OxygenInterpretation}

This case exemplifies situations when, in addition to information on the system's and the planets' properties, information is available that may indicate (or be consistent with) the presence of life. Such information would be broadly identified as a biosignature and may be an atmospheric biosignature, surface biosignature, technosignature, or any other type of information that may have similar significance. A specific example could be the simultaneous detection of atmospheric O$_2$ (at 9~$\sigma$-level) and CH$_4$ (at a 2$\sigma$-level) in the atmosphere of a planet with a radius of 1.3~$R_{\oplus}$ \citep[e.g.,][]{Sagan1993}.

While in Example 1 the primary goal is assessing the potential of a planet for habitability, in Example 2 the assessment of habitability informs the interpretation of a potential biosignature. If this assessment is performed, for example, through a Bayesian framework, then the assessment of habitability could be included as a prior.

In this case, we are making the assumption that planetary-scale oxygenic photosynthesis is more likely on a planetary surface that would provide a suitable habitat for terrestrial cyanobacteria. This point is important, as the temperature range in which cyanobacteria can grow is much more limited than the temperature range in which liquid water is stable or the temperature range in which Methanogens are viable (see Example 1 in \ref{S:Example1}).

\begin{figure}[ht!]
\centering
\includegraphics[width=\textwidth]{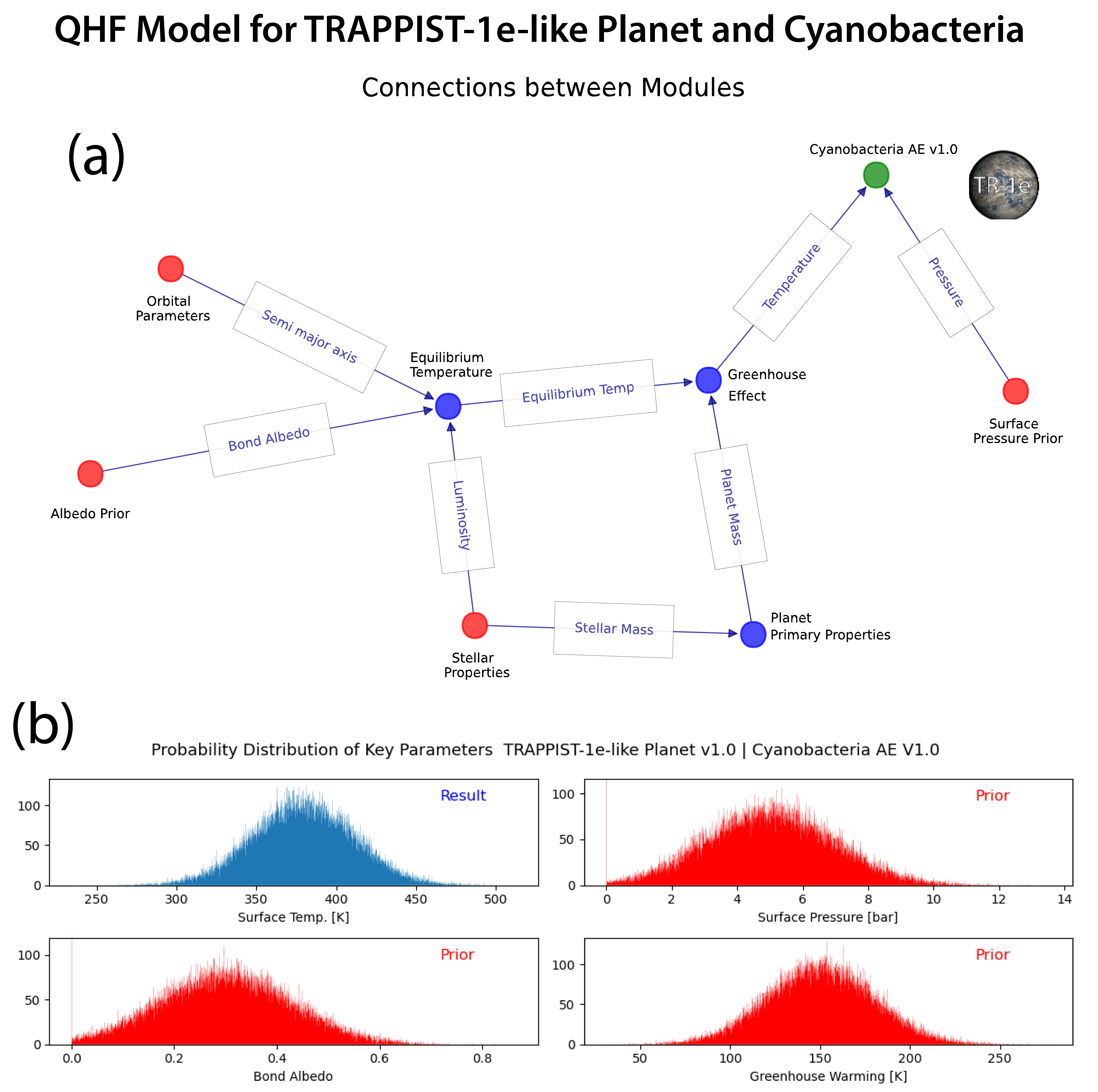}
\caption{QHF model assessment for a {\TRe{} planet} and cyanobacteria metabolism model, for Example 2. Panels a and b are the same as in Figure~\ref{f:Example1_TRAPPIST1e}.}
\end{figure}

\begin{figure}[ht!]
\centering
\includegraphics[width=\textwidth]{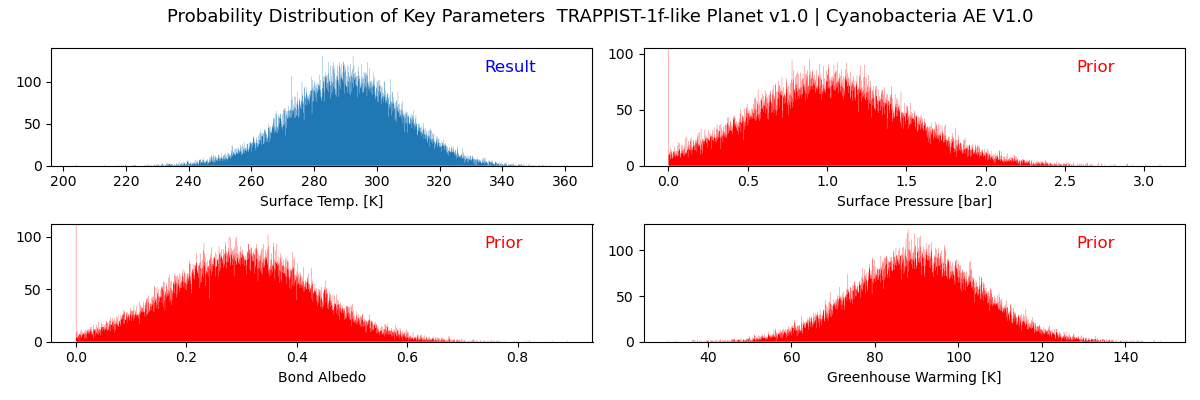}
\caption{Probability distributions for key prior and modeled parameters for a {\TRf{} planet} (based on the same module connections shown in Figure~\ref{f:Example1_TRAPPIST1e}). }
\end{figure}

We built a simple model for a {\TRe{} planet} and a {\TRf{} planet} and compared the model-predicted surface conditions to our model for the environmental requirements of cyanobacteria (see Section \ref{S:Cyanobacteria}). We use a Monte Carlo implementation of the \framework{} (described in more details in Section~\ref{S:Appendix}).

\begin{figure}[ht!]
\centering
\includegraphics[width=\textwidth]{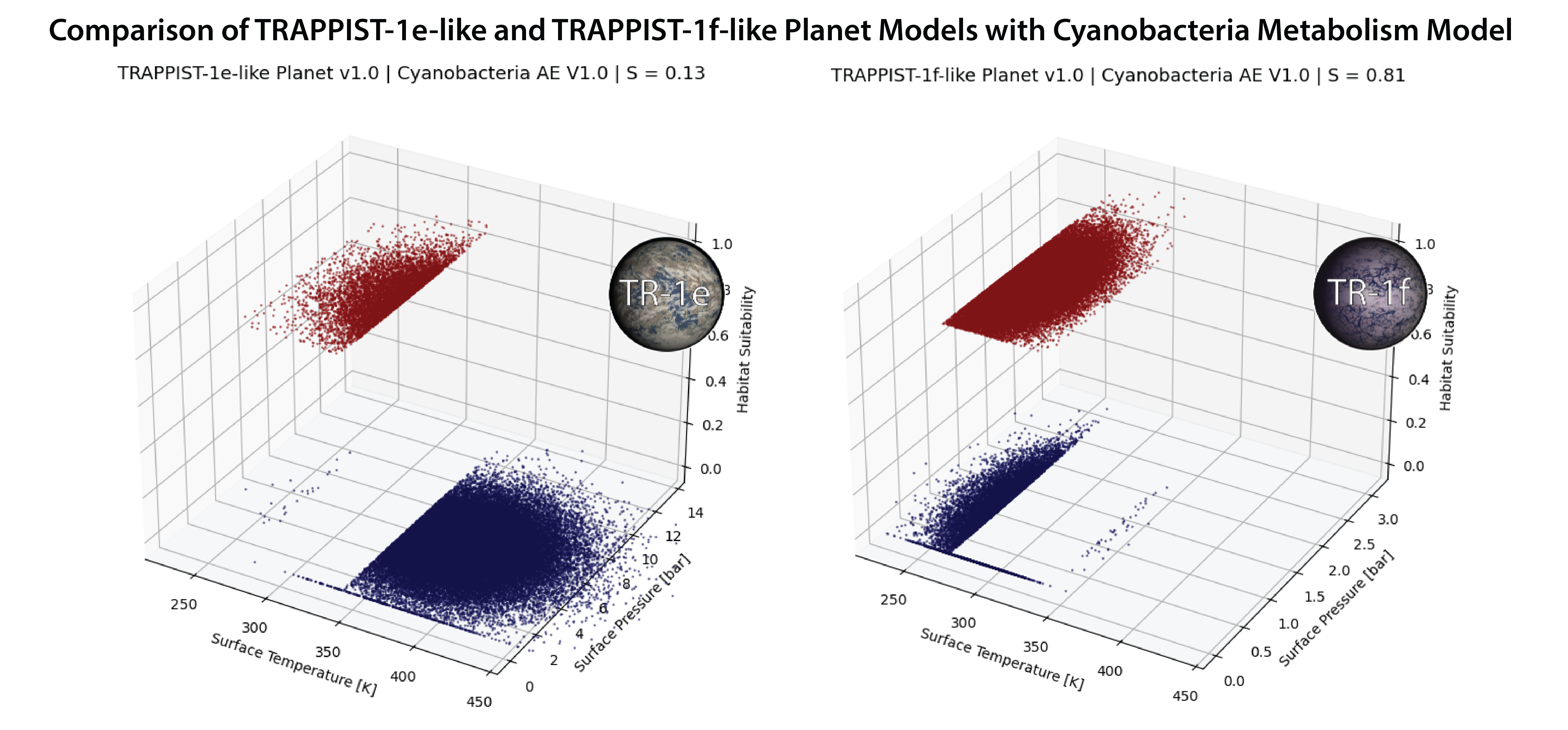}
\caption{Comparison of model-predictions for surface conditions {\TRe{} planet} and {\TRf{} planet} compared to the environmental conditions required by our Cyanobacteria model. Although both planets are likely to have surface conditions that allow stable liquid water, a {\TRe{} planet} would likely be too warm to allow Cyanobacteria (and terrestrial oxygenic photosynthesis) to exist. \label{f:Example2_Comparison}}
\end{figure}

In Figure~\ref{f:Example2_Comparison} we show the outcome of the assessment of two example habitat models for {\TRe{} planet} and a {\TRe{} planet} compared to the environmental requirements of cyanobacteria.

An interesting and non-trivial result emerges in Figure~\ref{f:Example2_Comparison}. Both planet models have surface conditions that are compatible with liquid water. However, due to its stronger greenhouse effect (in this particular model) {the surface of a \TRe{} planet} is warmer than that of {a \TRf{} planet}. The approximately 90~K temperature difference, when combined with the cyanobacteria viability function, offers a stark difference with respect to the assessment based on liquid water: The warmer {\TRe planet's} surface is likely too warm for cyanobacteria to be photosynthetically active and thus be the source of the oxygen detected in this example.
{The difference in the probability of habitat suitability demonstrates the utility of \framework{}: The interpretation of oxygen detection with or without considering the temperature range compatible with known oxygenic photosynthesis may lead to the opposite interpretation.}

\begin{deluxetable*}{c c c c}[t]
\tablecaption{Configurations and results from Examples 1--4. \label{T:ExampleResults}}
\tablehead{\colhead{\bf Example}  & \colhead{Habitat Model(s)} & \colhead{Metabolism Model} & \colhead{Habitat Suitability} }
\startdata
1            & \TRe   &  Archaea/Methanogens AE v1.0 &   $69\%$       \\
             & \TRf   &  Archaea/Methanogens AE v1.0& $93\%$         \\
\hline
2            & \TRe   &  Cyanobacteria AE v1.0&  $13\%$        \\
             & \TRf   &  Cyanobacteria AE v1.0&  $80\%$        \\
\hline
3            & Mars     &  Methanogens AE v1.0&   {Peak of $\sim 55\%$ at 5\,km depth}                \\
\hline
4            & Europa   &  Methanogens AE v1.0&  {Peak of $\sim 50\%$ at 42\,km depth} \\
\enddata
\end{deluxetable*}

\subsection{Example 3: Martian Surface and Sub-surface}
\label{S:ExampleMars}

In this example, we explore how a simple model for the Martian surface and sub-surface compares to the V-functions of model metabolisms. Our example aims to probe the primary effects impacting habitat suitability, but does not attempt to provide a realistic, comprehensive representation of the Martian environment. {We consider only the first-order environmental variables for the V-function (temperature $T$ and pressure)}. Our model for the Martian environment predicts the environmental parameters as a function of the elevation $z$, as follows: Temperature is calculated from a combination of a simple radiative energy balance equations and internal temperature gradient model (see Equation~\ref{Eq:MarsModel1}). For $z=0$ (i.e., surface), the temperature is set by the radiative energy balance. For $z<0$ (sub-surface), the heat from insolation becomes less important as the assumed geothermal heat is significant. {We assume that temperature will increase with depth following a temperature gradient ($K_t=\frac{dT}{dz}$)}. We assume a truncated ($K_t<0~K/m$), normal probability distribution. As the geothermal gradient {in the Martian subsurface is highly uncertain, we base our temperature distribution on the thermal gradient measured in Earth's lithosphere. This assumption is not unreasonable as a starting point given that typical heat flow values derived for Mars (e.g, 10--40 mWm/m$^{-2}$ \citealt[][]{Clifford2010} or 5--25 mWm/m$^{-2}$ \citealt[][]{Dehant2012}) are within the range of average continental heat flow rates measured for Earth (80$\pm$162 mWm$^{-2}$, e.g., \citealt[][]{Jaupart2007}).
Given this, we adopt $K_t=-0.03$~K/m (temperature increasing downward) as the mean value of the temperature gradient distribution (based on the lithospheric continental geotherm in Figure 12 of \citealt[][]{Jaupart2007}). To account for the broad range of heat flows measured in Earth's lithosphere and to account for likely differences in the Martian subsurface, we adopt a broad uncertainty range ($\sigma$=0.02 K/m).}
At the surface and above ($z\ge0$), the pressure is set by the atmospheric hydrostatic equilibrium. For the sub-surface, the pressure is calculated from the weight of the overlying column of regolith, assuming a typical constant density ($\rho =2,582$\,kg/m$^{3}$) and surface gravity (g=$3.73$\,m/s$^2$).
Water (either as liquid or ice) is presumed to be available in the sub-surface environment irrespective of the other parameters; but the availability of liquid water depends on the pressure/temperature conditions.
The model, thus, can be summarized as follows:

\begin{equation}\label{Eq:MarsModel1}
    T(Z) =
    \begin{cases}
        T_0 = \left( \frac{L_\odot (1-A_B)}{16 \pi a^2 \sigma \epsilon} \right)^{1/4} & \text{if  $z = 0$,}\\
        T_0 + z \times K_t  & \text{if  $z \leq 0$}
    \end{cases}
    \hskip 5mm
    p(z) =
    \begin{cases}
        p_0 \times e^{-z/H} & \text{if  $z > 0$,}\\
        p_0 - z \rho g  & \text{if  $z \leq 0$}
    \end{cases}
\end{equation}

\begin{figure}[ht!]
\begin{center}
\includegraphics[width=0.9\linewidth,angle=0]{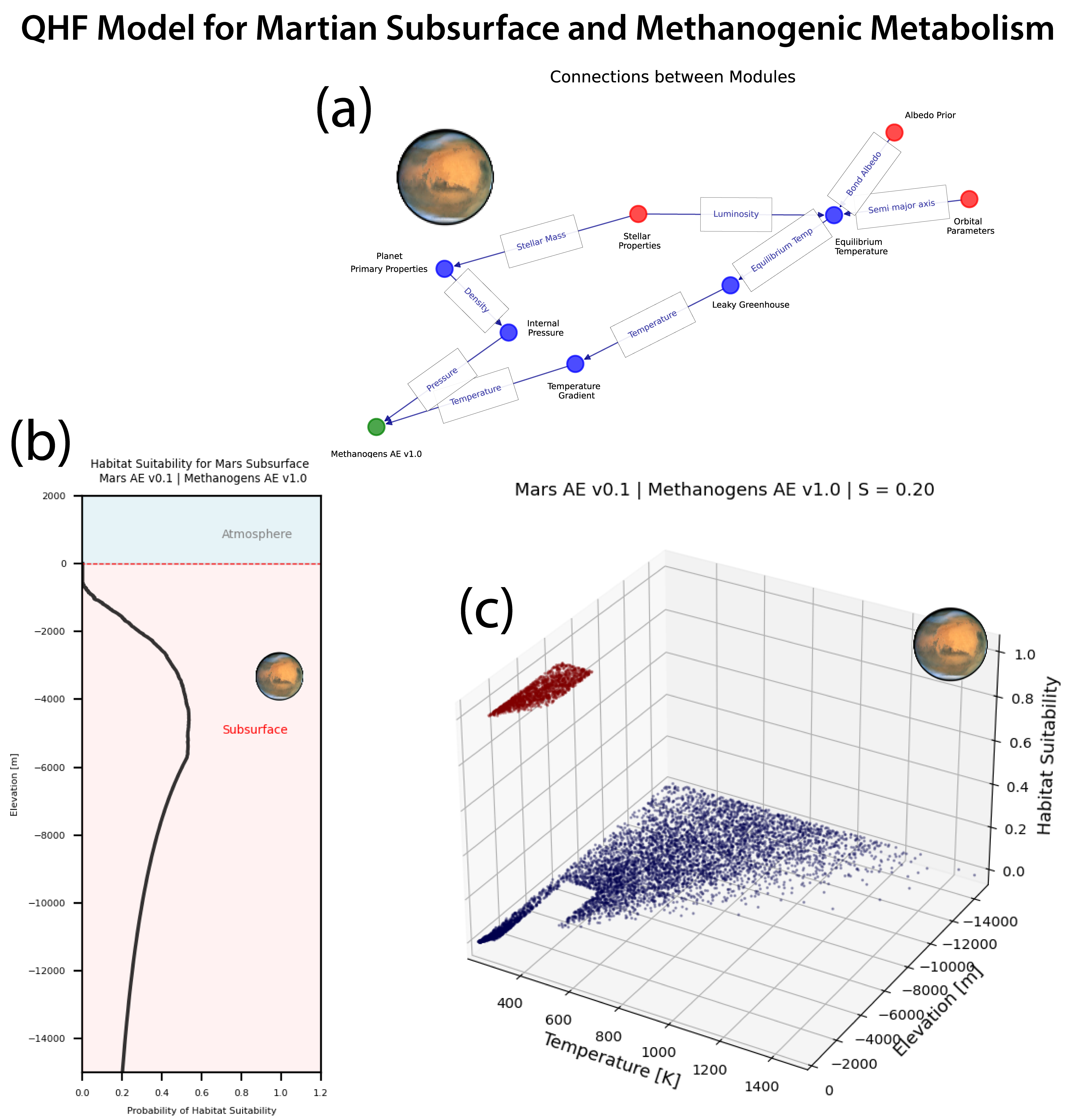}
\caption{Results of the habitat suitability assessment for the Martian subsurface with
Methanogens module. While temperatures are too low close to the surface, due to the geothermal heat gradient the deeper subsurface is warmer. It becomes more suitable for methanogens, up to the point where it begins the exceed the upper temperature limit of methanogen viability.  \label{f:MarsSubSurface}}
\end{center}
\end{figure}

Figure \ref{f:MarsSubSurface} summarizes the model and its results. Panel \textit{a} shows the graph of the model modules. Here, the priors are the Bond albedo, stellar properties, and orbital parameters. From these, the planet's fundamental properties, the surface pressure and surface temperature, are calculated through a leaky greenhouse model (see Section~\ref{S:Example1}). From the surface temperature and interior processes, the temperature and pressure conditions are calculated as a function of depth. Finally, the pressure and temperature conditions at different depths are compared to the viability function of the metabolism (see panel \textit{c}). The compatibility of the habitat model and the metabolism viability model results in the quantitative assessment of the habitat suitability.

Figure \ref{f:MarsSubSurface}c shows that -- in this simplistic Mars model -- the top {\textcolor{black}{$\sim$1~km of the subsurface is too cold to allow liquid water to be present. At greater depths ($\sim$1--8~km), however, the temperature is higher and liquid water becomes likely in the model, satisfying the conditions in our methanogen metabolism viability model. We find that the habitat suitability probability (Figure \ref{f:MarsSubSurface}b) peaks at a depth of approximately 5~km. However, at even greater depth ($>$9~km), the probability of temperatures being too high for methanogens begins to be significant and, though not shown in Figure~\ref{f:MarsSubSurface}, the sub-surface pressure also begins to exceed our assumed limit of 55~MPa for methanogen growth. This leads to lower and eventually zero habitat suitability.
}}

While our Mars model is admittedly simplistic, it demonstrates the use of multi-point evaluation of the \framework{}: We evaluated the habitat suitability for hundreds of points at different depth, thus providing a depth-dependent assessment. This use of the model is more complex than that demonstrated in Examples 1 and 2, and meets key requirements for use in the interpretation of in-situ studies in Solar System bodies.

\subsection{Example 4: Europa's Ocean}
\label{S:ApplicationEuropa}

In this example, we explore how compatible a simple model for the subsurface ocean in Europa is with the V-functions of model metabolisms introduced in Section~\ref{S:V-functions}. {As in the previous sub-section, our model is intended as an example and is constructed to be simple, capturing only primary effects.}

{We base our habitability assessment on a state-of-the-art geophysical model for Europa's ice layer, ocean, and interior. The one-dimensional internal structure model provides physically self-consistent solutions considering thermodynamics of aqueous solutions (with MgSO$_4$, NaCl, and NH$_3$) as well as pure water, and ice phases (\textsc{I}, \textsc{II}, \textsc{III}, \textsc{V}, and \textsc{VI}); silicates, and potential metallic core  \citep{Vance2018}. Furthermore, the structures we adopt also incorporate adjustments due to magnetic induction \citep{Vance2021}. Specifically, we adopt the temperature--depth profile from Figure~6 of \citet{Vance2023}, which are based on the two studies described above. These models predict a well-mixed, nearly isothermal ocean (with only 4--6\,K differential between the ocean floor and the ocean-ice interface) {with an average temperature gradient of approximately 0.025~$K/km$. We parametrize the temperature $T$ as a function of depth $d$ relative to the bottom of the ice layer as}}
\begin{equation}
T(d) = 1.48 \times 10^{-10} ~ \frac{K}{m^2}~ d^2 + 1.1 \times 10^{-5} ~ \frac{K}{m} ~ d + T_{\rm ice-water},
\end{equation}
{{where $T_{\rm ice-water}$ is the temperature of the ice--ocean transition at the bottom of the ice layer. Note that $d$ is a distance relative to the bottom of the ice layer so that $d$ increases as the absolute elevation $z$ (with respect to the surface) becomes increasingly negative. We assume the thickness of the ice layer to be 20 $\pm$ 4~km, and that the ocean terminates at a floor at $z = 125~km$.}}

\begin{figure}[ht!]
\begin{center}
\includegraphics[width=0.99\linewidth,angle=0]{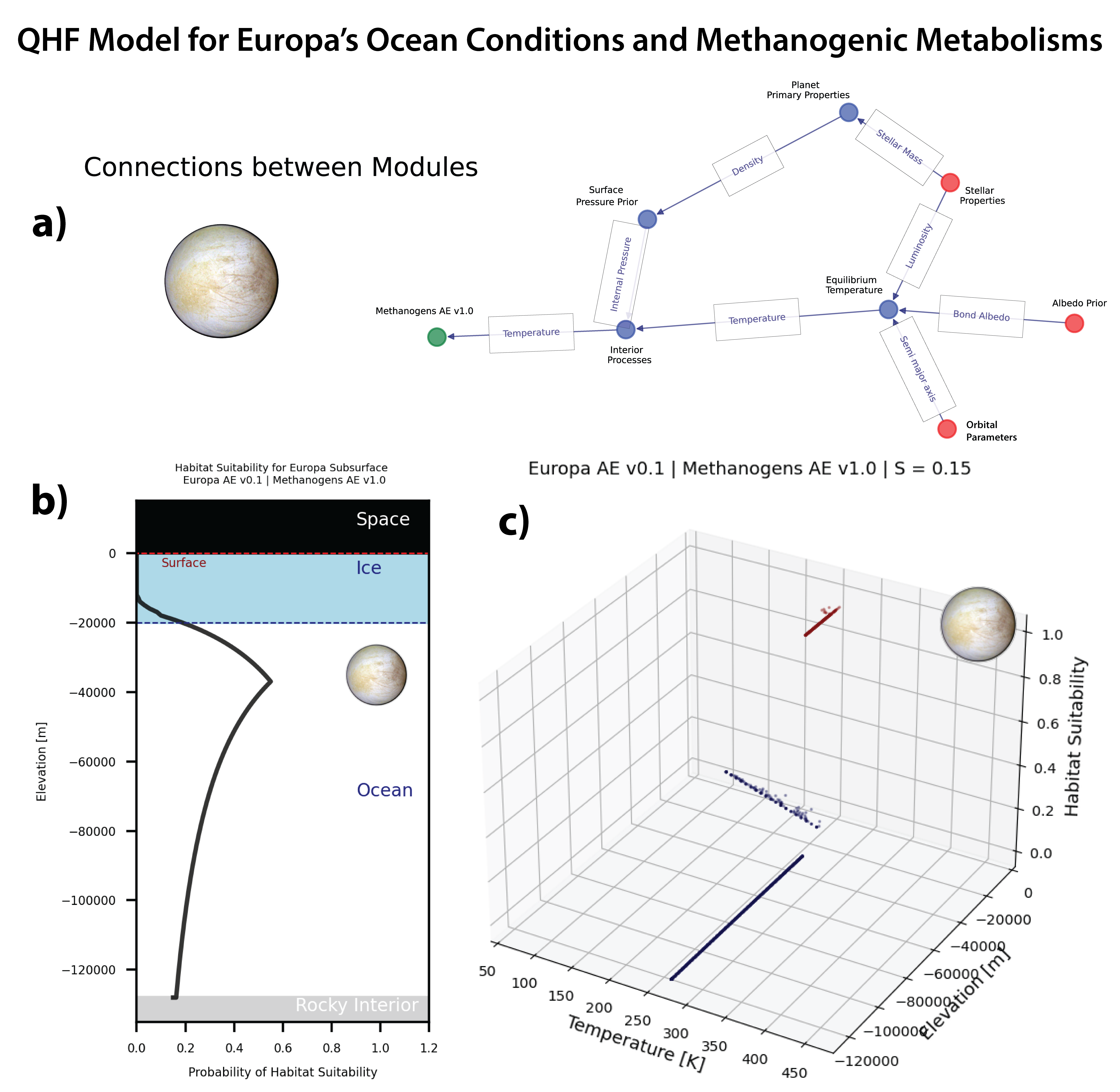}
\caption{Results of the habitat suitability assessment for Europa's subsurface ocean
considering the Methanogens module. {The ocean is nearly isothermal in our model \citep{Vance2018, Vance2021, Vance2023}, meaning the temperature is quite suitable for methanogens throughout its entirety. {However, though not shown in the figure, the pressure increases steadily with depth and ultimately limits the extent of suitability.}}  \label{f:EuropaSubSurface}}
\end{center}
\end{figure}

Similar to Figure~\ref{f:MarsSubSurface}, in Figure~\ref{f:EuropaSubSurface} we show an overview of the model structure and results of the habitat suitability assessment. Figure~\ref{f:EuropaSubSurface}a shows the representation of the model modules as a graph, with priors (orbital parameters, albedo, stellar parameters), and calculated parameters (radiative equilibrium temperature, surface temperature,  interior structure) and the final comparison of the probability distributions of the key state parameters (temperature and pressure) to the metabolism's viability function (methanogens AE v1.0).

Figure~\ref{f:EuropaSubSurface}b shows the graphical representation of the multi-layer model, and the result of the habitat suitability assessment as a function of depth. We note that the graphical representation of the multi-layer model shows an average ice thickness -- in our model, the uncertainty in the ice layer thickness is represented (see above) and sets the boundary condition for the ocean, and therefore {can} impact the pressure-temperature profile in the ocean. {In the example shown, we assume for simplicity that $T_{\rm ice-water}$ = 273.15~K but this could be customized to account for how additional variables like pressure and salinity affect this boundary region.} Figure~\ref{f:EuropaSubSurface}c shows the habitat suitability values as a function of surface temperature and depth. In this simplistic Europa subsurface model, {the ocean's nearly isothermal profile means that temperature is likely not the key limitation for the viability of methanogens. {Rather, the pressure is what sets the downward extent of suitability in our model. The pressure increases linearly from approximately 25~MPa (250~atm) at the ice-ocean boundary to 170~MPa (1675~atm) at the ocean floor, and intersects our assumed 55~MPa limit for methanogens at a depth of approximately 40~km. This suggests that, under our assumptions, Europa's upper ocean would be suitable for methanogens. If methanogens can indeed exist and grow at higher pressures \citep[e.g., the 75~MPa suggested by][]{Miller1988}, then a wider range of depths would be suitable. This example shows that that secondary variables can become a limiting factor in certain environments, which can be explored in more detail by incorporating more complex models of the ocean into our framework than we used here.}}

\section{Discussion}
\label{S:Discussion}

\subsection{New Terminology and Framework}

Given thorough consideration of the current state of the literature and a variety of approaches and needs, we introduced a new terminology for assessing the suitability of habitats for organisms. The new framework represents an extension of past terminology and methodology, and enables a more accurate and more transparent assessment of the compatibility of habitat-metabolism combinations.

The key differences between most past approaches and the new framework are as follows:
\begin{enumerate}
    \item Our assessment is based on the comparison of two models (habitat and metabolism), while most former assessments are based on the comparison of the habitat's derived properties and the stability conditions for liquid water.
    \item Our assessment is described as a model-dependent assessment, rather than the \textit{de facto} absolute assessments of many former studies.
    \item Our methodology is applied to and successfully demonstrated on multiple different science questions and environments, while most past approaches were specific to applications and not easily generalizable.
    \item Our assessment is provided with a modular, customizeable, open source software implementation
    that enables further refinements while maintaining cross-compatibility between studies.
\end{enumerate}

\subsection{Organisms, Species, Ecosystems, Metabolisms}
\label{S:Metabolism}

The \framework{} introduced here can be applied to a broad range of scales of living systems: From an individual cell, through a colony, a specific type of metabolism, to species, set of species, to ecosystems (from localized to even planetary-scale). This flexibility is provided by the highly customizable module that describes the viability requirements: Modules that correspond to different scales can be implemented with ease.

While from the mathematical and algorithmic perspective the model for the requirements of the living systems may in many cases be highly scalable or even scale-free, the common \textit{scientific terminology} used to describe living systems almost always implies a scale (e.g., cell, organism, cluster, colony, ecosystem, etc.).
We considered but decided against introducing an entirely new scale-free term for living systems.
Instead, we adopted the use of ``metabolism'' to refer to living systems of arbitrary scale but shared requirements for habitat conditions. This decision is motivated by the empirical fact that shared metabolism among species often correlates with environmental conditions in which the species can complete their lifecycles. Therefore, working with the term ``metabolism'' allows a more general grouping of types of living systems (regardless of scale) than other categories used in the literature.

We note that our framework could be applied to an ecosystem comprised of interacting species with different metabolisms. In such situations the metabolism model in our framework would correspond to the intersection of the sets of environmental conditions required by the species/metabolisms present.

\subsection{Example applications}

In this paper, we demonstrated the \framework{} on four example applications. These applications exemplified key use cases of the framework. While the individual examples and direct results are discussed in the relevant subsections (\S \ref{S:Example1}-- \ref{S:ApplicationEuropa}), we highlight here four over-arching properties of the framework that these examples demonstrate.

First, our framework is highly versatile: In the four examples, we applied it to extrasolar planets as well as to a Solar System rocky planet and a subsurface ocean on an icy moon. Furthermore, the purpose of the application varied through the four examples: In the first example (\S~\ref{S:Example1}), the framework facilitated the comparison of two exoplanets to help target prioritization. In the second example, the framework supported the interpretation of the detection of atmospheric biosignatures, considering both the planetary environments and terrestrial biological examples. In the third and fourth examples, the habitat suitability was explored in subsurface environments, which could guide mission objective definitions. These examples represent very different use cases and environments, and demonstrate how the \framework{} can be applied to them.

Second, the framework can incorporate and handle uncertainties in all key parameters and even in the model structure itself. Thanks to the modular Monte Carlo implementation, input parameters are efficiently drawn from probability distributions and then correctly propagated through the graph of the modules.

Third, the modular structure of the \framework{} enables coupling to (or integration) of models of different complexity. While our example models are kept as simple as possible, they demonstrate that more complex models (or outputs from more complex models) can be considered when evaluating the framework.

Fourth, the use of the \framework{} already led to some non-trivial results that would not have been possible with most previously used models. For example, in our second use case (\S~\ref{S:OxygenInterpretation}), we demonstrated how the framework can support the interpretation of a potential atmospheric biosignature. In this example, we assumed that atmospheric O$_2$ -- a potential biosignature -- is detected in two similar, but not identical exoplanets. The question is: In which of the two planets is the O$_2$ detection more likely to indicate genuine biosphere, i.e., the presence of large-scale oxygenic photosynthesis? {The majority of the extensive literature on the interpretation of oxygenic photosynthesis often considers possible sources and sinks (including atmospheric photochemistry and outgassing), and direct or indirect observations that can test the presence of water vapor in the atmosphere or (as liquid) on the planetary surface \citep[see, e.g.,][ and references therein]{Meadows2018}.}
{In this example, such} comparison would suggest that both planets are equally habitable, in spite of the moderate temperature difference between them. In contrast, the \framework{} shows an important difference: By comparing the conditions to those required by terrestrial cyanobacteria -- thought to be primary oxygen producers -- it becomes clear that the warmer planet is outside the temperature range in which cyanobacteria are photosynthetically active. Thus, by incorporating an admittedly simple metabolism model -- which is still more realistic than liquid water -- the conclusion of this comparison and the interpretation of the biosignature fundamentally changes. This example shows the gain from more detailed models that consider actual organisms rather than a phase of a solvent.

\subsection{Not a Universal Model but a Customizable, Modular Framework}

The examples for the framework introduced here are not intended to provide a complete universal assessment of habitat suitability. For instance, the simple energy balance model included for the exoplanet example is not intended to be the best possible model for that exoplanet.
Rather, the examples show how the framework can be \textit{customized} for specific cases. The framework provides a blueprint for how to connect models and assess joint likelihoods. It is envisioned that much more specific and comprehensive models for organisms and potential habitats will be integrated into (or connected) to the framework for individual applications.

For example, users may wish to integrate the VPlanet \citep[][]{Barnes2020} exoplanet evolution model suite (itself modular) to the framework presented here. Such a connection will combine comprehensive models for potential exoplanet habitat (from VPlanet) with the habitat suitability assessment from the \framework{} presented here.  Similarly, our framework could potentially be connected to PacMan \citep[][]{Krissansen-Totton2022} or PROTEUS \citep{2021JGRE..12606711L,2024JGRE..12908576N}, frameworks for modeling coupled atmosphere and interior evolution for rocky planets, or to PlaHab, a 2D energy balance model \citep[][]{Ramirez2024} that can simulate atmospheres of different compositions in rapidly and synchronously rotating planets.

\subsection{Model-dependent Assessment}

It is important to remember, that the habitability assessment resulting from the application of the \framework{} is not an absolute result, but a model-dependent inference. This is not a limitation, but an explicit recognition of the fact that our understanding of both the conditions in potential extraterrestrial habitats and of extraterrestrial life's requirements are based on sparse information. Therefore, model-based inferences are a necessity for comparing these; and, by extension, inferences on the compatibility of environmental conditions and necessary conditions for life will themselves be model-based.

To reflect this model-dependence, to ensure reproducability, scholars using the \framework{} should clearly identify the version of the framework, and the specific model for the metabolism, and for the habitat. Furthermore, we encourage authors to make the specific metabolism and habitat models used in their studies available for download.

\subsection{Contrasting with the Liquid Surface Water Habitability}

One example in our study highlights the advantage of the use of the metabolism viability functions over the liquid water conditions that past studies use as a stand-in for the temperature range terrestrial life can tolerate.

In our example (\S~\ref{S:OxygenInterpretation}) of the interpretation of an atmospheric biosignature (free O$_2$), we contrast the interpretation that would be reached when relying on liquid water stability range with the one that is reached when considering the temperature range in which cyanobacteria are viable. The liquid water temperature range (under 1 atm pressure) is a relatively broad range ($\sim 273 - 373$K). However, cyanobacteria are viable at a considerably narrower temperature range ($\sim273 - 343$~K). Consequently, in our example, if the biosignature is detected in a planet's atmosphere that is likely above $\sim350$~K, it is not likely to be compatible with the temperature range of cyanobacteria. This result is important, because cyanobacteria are widely considered to be the descendants of the only species in which oxygen photosynthesis has evolved. All other terrestrial organisms that utilize oxygenic photosynthesis emerged from endosymbiosis with cyanobacteria (e.g., chloroplasts in plants are derived from cyanobacteria).

Oxygenic photosynthesis is a highly complex process with multiple energy barriers that photosynthetic organisms overcome by dispensing energy generated by capturing visible light photons. It is reasonable to expect that oxygenic photosynthesis will universally require a highly complex chemical machinery. The fact that terrestrial oxygenic photosynthesis is limited to a relatively narrow temperature range is likely the consequence of the limited thermodynamical stability of the complex chemical pathways required.
Therefore, the detection of oxygen in an atmosphere of a planet too hot for cyanobacteria viability should \textit{not} be interpreted as a genuine biosignature, even if the planetary surface is compatible with the pressure/temperature conditions required by liquid water stability.

{The fact that oxygenic photosynthesis is limited to a narrower temperature range than the of water may seem obvious but, as we argue below, it is not commonly recognized in the exoplanet astrobiology community. The mission-level discussions in the exoplanet literature of the interpretation of oxygen detection do not consider this fact.
We illustrate this point by highlighting three influential examples:}

{Example 1: \citet{Lustig-Yaeger2019} entitled ``The Detectability and Characterization of the TRAPPIST-1 Exoplanet Atmospheres with JWST''.
This excellent, influential ($>$200 citations), 28-page-long comprehensive study explores the detectability of habitable environments in TRAPPIST-1 exoplanets; including biogenic O$_2$ and O$_3$. In this study, there is no mention of temperature ranges compatible with photosynthesis being potentially different from the temperature range required for an ocean.}

{Example 2: \citet{Lincowski2018} entitled ``Evolved Climates and Observational Discriminants for the TRAPPIST-1 Planetary System''. Another extensive (34 printed pages), very highly-cited ($>210$) study on detectability of atmospheric absorbers in TRAPPIST-1, including oxygen. This study does not consider or mention that the temperature range for oxygenic photosynthesis may not be identical to the temperature range for liquid water stability.}

{Example 3: LUVOIR Final Report \citep{TheLUVOIRTeam2019}: This superb report has fundamentally impacted NASA’s development of next-gen Flagship observatory. In the 700+ page report, we found no mention that the temperature range for photosynthetic organisms should be distinguished from the temperature range of liquid water stability. The specific discussion of O$_2$ and O$_3$ as ``Potential biosignatures'' (their Section 3.3.2) mentions the connection to liquid water but there is no mention of additional checks for an inferred temperature range being compatible with the temperature range for photosynthesis. The section on ``Confirmation of habitability'' describes the process of identifying habitable planets and interpreting biosignatures (including attempting to verify liquid water on the surface) and provides a specific sequence of steps and science questions (their Figure 3-11) that could be followed. Again, there is no mention of the temperature range for oxygenic photosynthesis being different from that required by liquid water.''}

{These three influential examples illustrate that -- while obvious to biologists -- most exoplanet-focused studies simply do not consider the basic habitat requirements beyond the temperature conditions for liquid water. (A further illustration of the point: The LUVOIR study mentions “oxygen” 37 times throughout the report but “photosynthesis” is mentioned only on 2 pages, always as a source that may create oxygen -- but not discussed or considered as a process.)}

{In contrast, QHF Example 2 demonstrates that such considerations are integral and quantitative components of our assessment process.}

\subsection{The Framework Separate from the Open-source Implementation}

We stress that the framework introduced here is not equivalent to the open source implementation. While the use of the python implementation \citep[][]{QHF} will be practical for many groups, some groups may want to adopt the framework but not the python implementation. These groups can do so: As long as the structure and numerical approach adopted conforms with the \framework{}, the results will be intercomparable to other studies. We encourage, however, these workers to retain the habitat and metabolism viability functions to ensure compatibility and minimize software divergence.

\subsection{Assessment in Evolving Systems}

Our \framework{} compares the conditions in a potential habitat with the requirements for metabolisms. The framework, as currently implemented, does not model time-varying systems; the examples provided assume steady-state systems. However, the functionality required for modeling time-evolving systems is part of the \framework{}: In Example 3, the depth-dependence of the Martian subsurface habitability is explored by assessing the habitat suitability in 500 distinct layers. The layers (and corresponding conditions) are parametrized through the use of ``probes'', essentially a variable that allows the repeated execution of the \framework{} at different spatial/temporal locations. Thus, that example can be easily modified and the ``probes'' can be deployed to model a time-evolving system rather than spatially extended system. If the habitat modules are time-dependent, the assessment can be executed at the desired points in time. For example, if the coupled time co-evolution of a habitat and a metabolism is modeled, then the habitat suitability at a representative times can be assessed with our framework.

\subsection{Habitat Suitability Assessment: From Empirical to Theoretical}

The examples included with the current paper for the use of the \framework{} all use parametric viability models based on empirical constraints on the environmental conditions in which known terrestrial metabolisms live. This \textit{empirical} basis for the examples presented here is a choice of convenience and not a general or necessary limitation of the framework. It is likely that, in the near future, metabolism viability models based on theoretical considerations will be developed. Such expansion will enable, for example, modeling metabolisms that are biochemically distinct from terrestrial life or living beyond the range of environmental conditions that known terrestrial life does. Given its modular, customizeable structure, our framework can be readily expanded with novel metabolism viability functions.

\subsection{Beyond Viability and Habitat Suitability}

Our framework's goal is to quantitatively assess the viability of a modeled metabolism in a modeled habitat. While this is an important assessment, in some situations -- where the habitat's properties are well understood -- it is possible to provide more comprehensive assessments of the ecosystems that may exist in it. Specifically, studies have successfully explored the viability of metabolisms and the biomass sustainable in the early Earth \citep[][]{Sauterey2020} and early Mars \citep[][]{Sauterey2022}. In other cases, such studies have been used -- via Bayesian assessments --- to interpret observational constraints on Enceladus' subsurface oceans and quantify a potential methanogenic ecosystem in it \citep[][]{Affholder2021}. We consider the quantitative habitability assessment presented in the current manuscript (best applicable to data-poor  potential habitats) as a precursor to ecosystem--biomass assessment (best applicable to data-rich habitats) presented in the above and similar studies.

\section{Summary}
\label{S:Summary}

In this paper, we provide an overview of approaches to develop habitability, a term with multiple, often contradictory uses, into a framework that is quantitative and applicable to a wide range of uses in the interdisciplinary context of astrobiology.

We surveyed the literature and identified key arguments for different uses and definitions of ``planetary habitability''. Careful reviews of published studies, committee reports, and specific examples helped us identify the set of most important considerations for habitability and its uses.

The following are the key points and findings of our study:

1) We provide a comprehensive overview of the context and uses of the terms ``habitability'' and ``habitable planets'', including the analysis of recent strategic reports. We also review relevant quantitative approaches and frameworks for assessing habitability.

2) We identify two opposing directions in which arguments in the literature would move habitability: Many scholars prefer simpler, narrow definitions that are more specific to Earth-like life on the surface of an Earth-like planet. In contrast, many workers prefer more universal definitions that are necessarily more complex and less specific.

3) At the very root of the contradictory uses and conflicting viewpoints on habitability lies the problem that we do not possess a universal definition for life: As we lack an understanding of the types of {non-terran} life that are possible, we are also unable to identify a set of requirements for life in general.

4) The ``process of elimination'' is commonly invoked to narrow down target samples and to identify the planets most likely to be ``habitable''. While this process is powerful, we show that due to mathematical asymmetry inherent to the problem, it cannot be used in reverse, i.e., probabilistically arguing for a location being ``habitable'' based on incomplete dataset (without evidence for the location being inhabited) is necessarily incorrect.

5) We propose a self-consistent, quantitative, modular, and customizable framework that builds on some past terminology but is distinct from those.

6) The \framework{} essentially consists of a probabilistic comparisons of \textit{two models}: A \textit{habitat model} for the potential habitat's properties (state vector), informed by prior knowledge and specific measurements; and by a \textit{metabolism viability model}
(``\textit{V-function}'') which describes the environmental conditions that are necessary for the robust viability of the modeled metabolism.

7) The \framework{} is highly customizable through the addition or expansion of the habitat modules and metabolism viability modules. In this initial publication, we include simple habitat examples for {\TRe} and {\TRf{} planets'}  surface conditions, Mars subsurface, and Europa's ocean. We also include example metabolism viability functions for liquid water, methanogenic archaea, and cyanobacteria.

8) Importantly, our framework can correctly handle uncertainties -- even if complex -- in priors and in the model description itself. The framework accounts for uncertainties through the use of the Monte Carlo sampling technique, where the habitat suitability assessment is repeated many times, each time sampling the priors' probability distributions randomly and repeating the chain of calculations through the module connections. From the ensemble of the hundreds of assessment runs, we express the probability of habitat suitability as the ratio of runs that lead to conditions matching viability criteria over the total number of runs. The implementation allows for the integration of highly complex models into the \framework{}, while correctly handling the uncertainties.

9) Our framework sets a standard and shareable format for the habitat and metabolism viability modules, which will enable the community to combine and compare results, and transfer solutions across research groups. Our framework can also readily interface with existing, complex planetary environment models.

10) In our framework, viability (\textit{V-functions}) are similar to habitability (H-functions) used in ecology \citep[e.g.,][]{Mendez2021}, but differ from H-functions in an important way: Their value reflects the compatibility of the environment with the metabolism, rather than the environment's carrying capacity. In other words, H-functions predict the carrying capacity of an environment, while V-functions do not.

11) We provide four example applications for the new framework. In our first example, we show how the framework can support the prioritization of candidate habitable exoplanets for biosignature searches. In our second example, we show how the framework can inform the interpretation of a potential atmospheric biosignature in an exoplanet atmospheres. In our third example, we explore the suitability of the Martian subsurface for methanogenic archaea. In our fourth example, we study the habitability of Europa's oceans for photosynthetic algae (cyanobacteria) and for methanogens.

12) We provide a python implementation of the \framework{} \citep[][]{QHF}. We include example applications as presented in this paper. This open source framework allows the community to further develop the framework and to apply it to a multitude of situations.

Our study provides a reference framework for quantitatively assessing the potential viability of various metabolisms in different habitats. An open-source implementation can be easily adopted for specific problems; and, vice versa, existing, comprehensive modeling frameworks can be adjusted to conform to the \qfw{} standards. Our work provides a basis for a uniform, quantitative assessment of habitability, a core concept that underpins the next decades of research in astrobiology and extrasolar planets.

\vskip 3mm
\textbf{Acknowledgements:} This material is based upon work supported by the National Aeronautics and Space Administration under Agreement No. 80NSSC21K0593 for the program “Alien Earths”. The results reported herein benefited from collaborations and/or information exchange within NASA’s Nexus for Exoplanet System Science (NExSS) research coordination network sponsored by NASA’s Science Mission Directorate. RB acknowledges support from NASA grants 80NSSC20K0229 and 80NSSC18K0829. Discussions with Charles Cockell, Jamie Dietrich, Shawn Domagal-Goldman, Vikki Meadows, Aki Roberge, Avi Mandell, Nancy Kiang helped clarify the context, uses, and considerations for the quantitative habitability framework. We acknowledge the NExSS Quantitative Habitability Science Working Group and its members for their active participation in the many discussions that helped to shape this study.
We thank the two anonymous referees whose comments greatly enhanced the quality and clarity of our manuscript.

\vskip 3mm
\textbf{Author Contributions:} DA proposed the Quantitative Habitability Science Working Group (SWG), led its meetings for two years, including the workshop. DA led the consolidation of different perspectives into a new framework, the development of the manuscript, and wrote the python implementation of the framework. RB co-chaired the SWG between 2022 and 2024, contributed significantly to the development of the framework, contributed to the definition of the examples, and contributed to the manuscript. MM summarized the relevant major reports, led the revision of the initial Europa ocean model, and made major contributions to the python framework.  TL made important contributions and ideas to the development of the framework and to the manuscript. NT contributed to the definition of the framework  and made substantial contributions to the manuscript. RF, AMe, AA, provided important perspectives on connecting habitability of extrasolar environments to methodology used in terrestrial ecology. SK actively participated in the SWG meetings and contributed to the definition of the framework, the historical and scientific context for habitability, and contributed to the manuscript. RM provided comments on the manuscript draft and helped refine its scope. VK contributed to changes in the python framework and to definitions of the metabolism models.

\pagebreak

\appendix

\section{Open Source Implementation}
\label{S:Appendix}

In order to facilitate the use and development of the framework described in this work, we provide a Python-based example implementation \citep[][]{QHF}. Newer versions of the python framework are available via a GitHub repository\footnote{https://github.com/danielapai/QHF}. Key figures in this manuscript (e.g., Figures~\ref{f:Example1_TRAPPIST1e}, \ref{f:Example1_TRAPPIST1f}) were prepared using this software framework.

Our Python example implementation uses standard libraries and demonstrates the formulation of the assessment, including the definition of the modules and connections (or, in a graph, nodes and edges). Detailed instructions are provided in the library (see the README file) and we only briefly describe the general structure and functionality of the program.

\begin{figure}[ht!]
\centering
    \includegraphics[width=0.45\textwidth]{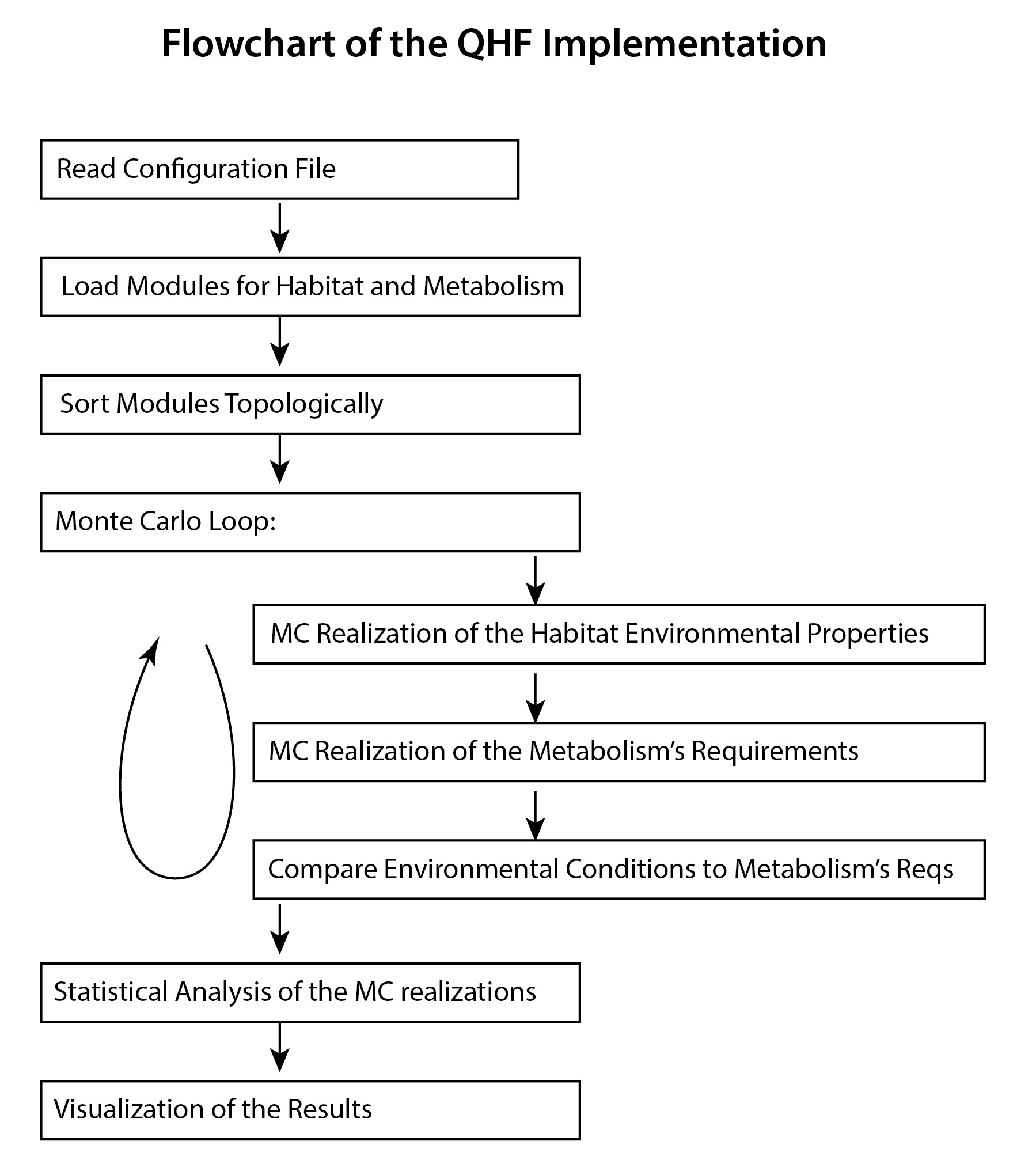}
    \caption{Flowchart of the Python implementation of the \qfw{} presented in this paper. The biological and physical models are imported as modules, which are then sorted into an acyclical graph and evaluated many times in a Monte Carlo approach. The functionality of the framework is configured through the modules included, providing a flexibility and upgradeability to the framework.\label{F:ImplementationFlowchart}}
\end{figure}

Figure~\ref{F:ImplementationFlowchart} shows a flowchart-style representation of the high-level organization of the open source implementation provided with this paper.

\bibliography{quanthabrefs}{}
\bibliographystyle{aasjournal}

\end{document}